\documentclass[12pt]{article}

\usepackage{ametsoc}
\usepackage{amsmath}

\usepackage{graphicx}
\usepackage{epsfig}
\usepackage{fullpage}
\usepackage{setspace}

\newcommand{\p}{{\bf p}}

\begin{document}

\title{\bf Resonant and Near-Resonant Internal Wave Interactions}

\author{Yuri V. Lvov$^1$, Kurt L. Polzin$^2$ and Naoto Yokoyama$^3$\\
{\small $^1$ Department of Mathematical Sciences, Rensselaer
Polytechnic Institute, Troy NY 12180}\\
{\small $^2$ Woods Hole Oceanographic Institution, MS\#21, Woods Hole, MA 02543}\\
{\small $^3$ Department of Aeronautics and Astronautics, Kyoto University, Kyoto, Kyoto 606-8501 JAPAN}
}

%
%

\amstitle

\begin{abstract}

We report evaluations of a resonant kinetic equation that suggest the slow time evolution of the Garrett and Munk spectrum is {\em not}, in fact, slow.  Instead nonlinear transfers lead to evolution time scales that are smaller than one wave period at high vertical wavenumber.   Such values of the transfer rates are inconsistent with conventional wisdom that regards the Garrett and Munk spectrum as an approximate stationary state and puts the self-consistency of a resonant kinetic equation at a serious risk.  We explore possible reasons for and resolutions of this paradox.  

Inclusion of near-resonant interactions decreases the rate at which the spectrum evolves.  This leads to improved self-consistency of the kinetic equation.

\end{abstract}

\clearpage
\section{Introduction}  \label{Introduction}

Wave-wave interactions in stratified oceanic flows have been a subject of intensive research in the last four  decades.  Of particular importance is the existence of a ``universal'' internal-wave spectrum, the Garrett and Munk spectrum.  It is generally perceived that the existence of a universal spectrum is, at least in part and perhaps even primarily, the result of nonlinear interactions of waves with different wavenumbers.  Due to the quadratic nonlinearity of the underlying primitive equations and the fact that the linear internal-wave dispersion relation can satisfy a three-wave resonance condition, waves interact
in triads.  Therefore the question arises: how strongly do waves within a given triad interact?  What are the oceanographic consequences of this interaction?

Wave-wave interactions can be rigorously characterized by deriving a closed equation representing the slow time evolution of the wavefield's wave action spectrum.  Such an equation is called a {\em kinetic equation} \citep{ZLF} and significant efforts in this regard are listed in Table \ref{TABLEOFELEMENTS2}.

A kinetic equation describes, under the assumption of weak nonlinearity, the resonant spectral energy transfer on the {\it resonant manifold}.  The resonant manifold is a set of wavevectors ${\bf p}$,  ${\bf p_1}$ and ${\bf p_2}$ that satisfy 
\begin{equation} \p = \p_1 + \p_2, \ \ \ 
\omega_{\p} = \omega_{\p_1}+ \omega_{\p_2}, \label{ResonantManifold}
\end{equation}
where the frequency $\omega$ is given by a linear dispersion relation relating wave frequency $\omega$ with wavenumber $\bf{p}$.  

The reduction of all possible interactions between three wavevectors to a resonant manifold is a significant simplification. {\it Even further} simplification can be achieved by taking into account that, of all interactions {\it on} the resonant manifold, the most important are those which involve extreme scale separations~\cite{MB77} between interaction wavevectors. It is shown in ~\cite{McComas} that Garrett and Munk spectrum of internal waves is stationary with respect to one class of such interactions, called Induced Diffusion.  Furthermore, a comprehensive inertial-range theory with constant downscale transfer of energy was obtained by patching these mechanisms together in a solution that closely mimics the empirical universal spectrum (GM)\citep{MM81b}.  It was therefore concluded that that Garrett and Munk spectrum constitutes an approximate stationary state of the kinetic equation.  

In this paper we revisit the question of relation between Garrett and Munk spectrum and the resonant kinetic equation.  At the heart of this paper (Section \ref{resonant_time}) are numerical evaluations of the \cite{LT2} internal wave kinetic equation demonstrating changes in spectral amplitude at a rate less than an inverse wave period at high vertical wavenumber for the Garrett and Munk spectrum.  This rapid temporal evolution implies that the GM spectrum is {\it not} a stationary state and is contrary to the characterization of the GM spectrum as an inertial subrange.  This result gave us cause to review published work concerning wave-wave interactions and compare results.  The product of this work is presented in Sections \ref{ResonantInteractions}$\&$\ref{Rotations}.  In particular, we concentrate on four different versions of the internal wave kinetic equation:  
\begin{itemize}
\item a noncanonical description using Lagrangian coordinates \citep{O74,O76,MO75},
\item  a canonical Hamiltonian description in Eulerian coordinates \citep{Voronovich},
\item a dynamical derivation of a kinetic equation without use of Hamiltonian formalisms in Eulerian coordinates \citep{Zeitlin},
\item a canonical Hamiltonian description in isopycnal coordinates \citep{LT,LT2}.
\end{itemize}
We show in Section \ref{ResonantInteractions} that, without background rotation, all the listed approaches are {\it equivalent} on the resonant manifold.  In Section \ref{Rotations}  we demonstrate that the two versions of the kinetic equation that consider non-zero rotation rates are again {\it equivalent} on the resonant manifold.  This presents us with our first paradox:  if all these kinetic equations are the same on the resonant manifold and exhibit a rapid temporal evolution, then why is GM considered to be a stationary state?  The resolution of this paradox, presented in Section \ref{discussion}, is that:  (i) 
numerical evaluations of the \cite{McComas} kinetic equation demonstrating the induced diffusion stationary states require damping in order to balance the fast temporal evolution at high vertical wavenumber, and (ii) the high wavenumber temporal evolution of the \cite{LT2} kinetic equation is tentatively identified as being associated with the elastic scattering mechanism rather than induced diffusion.  

Having clarified this, we proceed to the following observation:  Not only do our numerical evaluations imply that the GM spectrum is {\it not} a stationary state, the rapid evolution rates correspond to a strongly nonlinear system.  Consequently the self-consistency of the kinetic equation, which is built on an assumption of weak nonlinearity, is at risk.  Moreover, reduction of all {\em resonant} wave-wave interactions exclusively to extreme scale separations is also not self-consistent.  

Yet, we are not willing to give up on the kinetic equation. Our second paradox is that, in a companion paper \citep{Theory} we show how a comprehensive theory built on a scale invariant {\it resonant} kinetic equation helps to interpret the {\it observed variability} of the background oceanic internal wavefield.  The observed variability, in turn, is largely consistent with the induced diffusion mechanism being a stationary state{\bf !}  

Thus the resonant kinetic equation demonstrates promising predictive ability and it is therefore tempting to move towards a self-consistent internal wave turbulence theory.  One possible route towards such theory is to include to the kinetic equation near-resonant interactions, defined as

\begin{equation} 
\p = \p_1 + \p_2, \nonumber \\ 
\mid \omega_{\p} - \omega_{\p_1}- \omega_{\p_2} \mid<\Gamma, 
\label{NearResonantManifold}
\end{equation}
where $\Gamma$ is the resonance width. We show in Section \ref{near_resonant_time} that such resonant broadening leads to slower evolution rates, potentially leading to a more self consistent description of internal waves. 

We conclude and list open questions in Section \ref{Conclusion}.  Our numerical scheme for evaluating near-resonant interactions is discussed in Section \ref{RBnum}.  An appendix contains the interaction matrices used in this study.

\section{Background}\label{Background}

A kinetic equation is a closed equation for the time evolution of the wave action spectrum in a system of weakly interacting waves.  It is usually derived as a central result of wave turbulence theory.  The concepts of wave turbulence theory provide a fairly general framework for studying the statistical steady states in a large class of weakly interacting and weakly nonlinear many-body or many-wave systems.  In its essence, classical wave turbulence theory~\citep{ZLF} is a perturbation expansion in the amplitude of the nonlinearity, yielding, at the leading order, linear waves, with amplitudes slowly modulated at higher orders by resonant nonlinear interactions.  This modulation leads to a redistribution of the spectral energy density among space- and time-scales. 

While the route to deriving the spectral evolution equation from wave amplitude is fairly standardized (Section \ref{B2}), there are substantive differences in obtaining expressions for the evolution equations of wave amplitude $a$.  Section \ref{A1} describes various attempts to do so.

\subsection{Hamiltonian Structures and Field Variables}\label{A1}
\subsubsection{A canonical Hamiltonian formulation in isopycnal coordinates}
 \cite{LT,LT2} start from the primitive equations of motion written in isopycnal coordinates:
\begin{eqnarray}
\frac{\partial}{\partial t}\frac{\partial z}{\partial \rho} + \nabla \cdot \left(\frac{\partial z}{\partial \rho} {\bf u} \right) &=& 0 , \nonumber \\
\frac{\partial {\bf u}}{\partial t} +f {\bf u}^\perp+ {\bf u} \cdot
\nabla {\bf u} + \frac{\nabla M}{\rho_0} &=& 0 ,
\nonumber
\\
\frac{\partial M}{\partial \rho} - g z &=& 0 .
\label{PrimitiveEquations}
\end{eqnarray}
representing mass conservation, horizontal momentum conservation under the Bousinesq approximation and hydrostatic balance.
The velocity ${\bf u}$ is then represented as: 
\begin{equation}
{\bf u} = \nabla \phi + \nabla^{\perp} \psi, \nonumber
\end{equation}
with $\nabla^{\perp}=(-\partial / \partial y, \partial / \partial x)$ 
and a normalized differential layer thickness is introduced:  
\begin{equation}
\Pi = \rho / g \partial^2 M/\partial \rho^2 = \rho \partial z/\partial \rho 
\end{equation}
Since both potential vorticity and density are conserved along particle trajectories, an initial profile of the potential vorticity that is a function of the density will be preserved by the flow.  
%
%
Hence it is self-consistent to assume that the potential vorticity $q$ is function of $\rho$ only, independent of $x$ and $y$:
\begin{equation}
  q(\rho) = q_0(\rho) = \frac{f}{\Pi_0(\rho)} \, ,
 \label{PV}
\end{equation}
where $\Pi_0(\rho) = -g / N(\rho)^2$ is a reference stratification profile with background buoyancy frequency, $N = (-g/(\rho \partial z/\partial\rho|_{\mathrm{bg}}))^{1/2}$, independent of $x$ and $y$. 
The  variable $\psi$ can then be  eliminated by assuming that potential vorticity is constant on an isopycnal so that $f + \Delta \psi = q_0 \Pi$ and one obtains two equations in $\Pi$ and $\phi$:  
\begin{eqnarray}
\Pi_t + \nabla \cdot (\Pi (\nabla \phi + \nabla^{\perp} \Delta^{-1} (q_0 \Pi -1)))  & = 0 \nonumber \\
\phi_t + \frac{1}{2} \mid \nabla \phi + \nabla^{\perp}\Delta^{-1}(q_0\Pi- 1)  \mid^2 + 
\Delta^{-1}\nabla \cdot [q_0 \Pi (\nabla^{\perp}\phi - \nabla \Delta^{-1}(q_0 \Pi - 1)) ] + & \nonumber \\
\frac{1}{\rho} \int^{\rho} \int^{\rho^{\prime}} \frac{\Pi - \Pi_0}{\rho_{1}} d\rho_1 d\rho^{\prime} & = & 0 
\end{eqnarray}
Here $\Delta^{-1}$ is the inverse Laplacian and $\rho^{\prime}$ represents a variable of integration rather than perturbation.  Serendipitously, the variable $\Pi$ is the canonical conjugate of $\phi$: 
\begin{equation}
 \frac{\partial \Pi}{\partial t}  = \frac{\delta {\cal H}}{\delta \phi} \, ,
  \qquad
\frac{\partial \phi}{\partial t}  = - \frac{\delta {\cal H}}{\delta \Pi} \, ,
  \label{Canonical}
\end{equation}
under a Hamiltonian ${\cal H}$:
%
\begin{eqnarray}
 {\cal H} = \!\! \int \!\! d {\bf x} d \rho
 \left(
 - \frac{1}{2} \left(
\Pi_0+\Pi({\bf x}, \rho) \right) \,
  \left|\nabla \phi({\bf x}, \rho) + \frac{f}{\Pi_0} \nabla^{{\perp}}\Delta^{-1} \Pi({\bf x}, \rho)
  \right|^2
+
  \frac{g}{2} \left|\int^{\rho} d\rho^{\prime} \frac{\Pi({\bf x}, \rho^{\prime})}{\rho^{\prime}} \right|^2
  \right) .
\nonumber\\
  \label{HTL2}
\end{eqnarray}
%
that is the sum of kinetic and potential energies.   

Switching to Fourier space, and
introducing a complex field variable $a_{{\bf p}}$ through the transformation
\begin{eqnarray}
 \phi_{{\bf p}} &=& \frac{i N \sqrt{\omega_{{\bf p}}}}{\sqrt{2 g} |{\bf k}|} \left(a_{{\bf p}}-
a^{\ast}_{-{{\bf p}}}\right)\, ,\nonumber\\
\Pi_{{\bf p}} &=& \Pi_0-\frac{N\, \Pi_0\,
|{\bf k}|}{\sqrt{2\, g \omega_{{\bf p}}}}\left(a_{{\bf p}}+a^{\ast}_{-{{\bf p}}}\right)\, ,
\label{transformationToSingleEquation2}
\end{eqnarray}
where the frequency $\omega$ satisfies the linear dispersion relation 
\begin{eqnarray}
\omega_{{\bf p}}=\sqrt{ f^2 + \frac{g^2}{\rho_0^2 N^2} \frac{|{\bf k}|^2 }{m^2}}~~,
\label{InternalWavesDispersion}
\end{eqnarray}
the equations of motion
(\ref{PrimitiveEquations}) adopt the canonical form
\begin{equation}
i\frac{\partial}{\partial t} a_{{\bf p}} = \frac{\delta {\cal H}}{\delta a_{\p}^{\ast}} ~ ,
\label{fieldequation}
\end{equation}
with the Hamiltonian
\begin{eqnarray}
&& {\cal H} = \int d{\bf p} \, \omega_{{\bf p}} |a_{{\bf p}}|^2
\nonumber \\
 && 
\quad
+ \int d{\bf p}_{012}
\left(
 \delta_{{\bf p}+{\bf p}_1+{\bf p}_2} (U_{{\bf p},{\bf p}_1,{\bf p}_2} a_{{\bf p}}^{\ast} a_{{\bf p}_1}^{\ast} a_{{\bf p}_2}^{\ast} + \mathrm{c.c.})
+ \delta_{-{\bf p}+{\bf p}_1+{\bf p}_2} (V_{{\bf p}_1,{\bf p}_2}^{{\bf p}} a_{{\bf p}}^{\ast} a_{{\bf p}_1} a_{{\bf p}_2} + \mathrm{c.c.}) ~.
\right)\nonumber\\
\label{HAM}
\end{eqnarray}

Eq. (\ref{fieldequation}) is Hamilton's equation and (\ref{HAM}) is the standard form of the Hamiltonian of a system dominated by three-wave interactions~\citep{ZLF}.
Calculations of interaction coefficients $U$ and $V$ are tedious but straightforward task, completed in \cite{LT,LT2}.

We emphasize that  (\ref{fieldequation}) is, with simply a Fourier decomposition and assumption of uniform potential vorticity on an isopycnal,  {\it precisely equivalent} to the fully
nonlinear equations of motion in isopycnal coordinates (\ref{PrimitiveEquations}). 
All other formulations of an internal wave kinetic equation depend upon a linearization prior to the derivation of the kinetic equation via an assumption of weak nonlinearity.  

The difficulty is that, in order to utilize Hamilton's Equation (\ref{fieldequation}), the Hamiltonian (\ref{HTL2}) must {\it first} be constructed as a function of the generalized coordinates and momenta ($\Pi$ and $\phi$ here).  It is not always possible to do so {\em directly}, in which case one must set up the associated Lagrangian (${\cal L}$ below) and then calculate the generalized coordinates and momenta.  

\subsubsection{Hamiltonian formalism in Clebsch variables in \citep{Voronovich}}

Voronovich starts from the non-rotating equations in Eulerian coordinates: 
\begin{eqnarray}
\frac{\partial {\bf u}}{\partial t} + {\bf u} \cdot \nabla {\bf u} & = & \frac{-1}{\rho} \nabla p  - g\hat{z} \nonumber \\
\nabla \cdot {\bf u} = 0 \nonumber \\
\frac{\partial \rho }{\partial t} + {\bf u} \cdot \nabla \rho & = & 0 ~ . 
\end{eqnarray}

The Hamiltonian of the system is
\begin{eqnarray}
{\cal H} = \int \left( (\rho_0+\rho)
\frac{{\bf v}^2}{2} + \Pi(\rho_o + \rho) -\Pi(\rho_o) + \rho g z
\right)
d {\bf r},
\label{VoronovichH}
\end{eqnarray}
where $\rho_0(z)$ is the equilibrium density profile, $\rho$ is the wave perturbation and $\Pi$ is a potential energy density function:  
\begin{equation}
\Pi(\rho_o + \rho) -\Pi(\rho_o) + \rho g z = g \int_{\eta (\rho_o+\rho)}^{\eta (\rho_o)} [\rho_0+\rho-\rho_0(\xi)] d\xi
\label{VsPEfunction}
\end{equation}
with $\eta(\xi)$ being the inverse of $\rho_o(z)$.  The intent is to use $\rho$ and Lagrange multiplier $\lambda$ as the canonically conjugated Hamiltonian pair:  
\begin{eqnarray}
\dot{\lambda} = \frac{\partial \cal{ H}}{\partial \rho} & = & - ({\bf v} \nabla ) \lambda + g(z-\eta(\rho_o+\rho))\\
\dot{\rho} = -\frac{\partial \cal{ H}}{\partial \lambda} & = & -({\bf v} \nabla)(\rho_o + \rho) \nonumber\\ 
\label{VoronovichHamPair}
\end{eqnarray}
with $z-\eta(\rho_o+\rho)$ being the vertical displacement of a fluid parcel and the second equation representing continuity.  The issue is to express the velocity ${\bf v}$ as a function of $\lambda$ and $\rho$, and to this end one introduces yet another function $\Phi$ with the harmonious feature 
\begin{equation}
\frac{\delta {\cal H}}{\delta \Phi} = 0
\end{equation} 
and a constraint.  That constraint is provided by: 
\begin{equation}
\nabla \cdot {\bf v} = - \frac{ \delta {\cal H} }{\delta \Phi} = 0 
\label{constraint}
\end{equation}
\cite{Voronovich} then identifies the functional relationship:  
\begin{equation}
{\bf v} = \frac{1}{\rho_0+\rho} \left( \nabla \Phi + \lambda\nabla(\rho_0+\rho)\right) \cong  \frac{1}{\overline{\rho}}\left(\nabla \Phi + \lambda\nabla(\rho_0+\rho) \right),
\label{VoronovichV}
\end{equation}
with the right-hand-side representing the Boussinesq approximation.  The only thing stopping progress at this point is the explicit appearance of $\xi$ in (\ref{VsPEfunction}), and to eliminate this explicit dependence a Taylor series in density perturbation $\rho$ relative to $\rho_0$ is used to express
the potential energy in terms of $\rho$ and $\lambda$.  The resulting Hamiltonian $\cal{ H}$ is
\begin{equation}\displaystyle
{\cal H} = \int [\frac{v^2}{2} + \Pi(\rho_o + \rho) -\Pi(\rho_o) + \rho g z] d{\bf r} \cong \frac{1}{2} \int [\lambda \nabla (\rho_o + \rho) (\nabla \Phi + \lambda \nabla(\rho_o+\rho)) -\frac{g}{\rho_o^{\prime}}\rho^2 + \frac{g \rho_o^{\prime \prime}}{\rho_o^{\prime 3}} \frac{\rho^3}{3}] d{\bf r} \label{VoronovichHamFinal}
 \end{equation}
with primes indicating $\partial / \partial z$.

The only approximations that have been made to obtain
(\ref{VoronovichHamFinal}) are the Bousinesq approximation in the
nonrotating limit, the specification that the velocity be represented
as (\ref{VoronovichV}) and a Taylor series expansion.  The Taylor series expansion is used to 
express the Hamiltonian in terms of canonically
conjugated variables $\rho$ and $\lambda$.  Truncation of this Taylor series is the essence of the slowly varying (WKB) approximation that the vertical scale of the internal wave is smaller than the vertical scale of the background stratification, which requires, for consistency sake, the hydrostatic approximation.  

The procedure of introducing additional functionals ($\Phi$) and constraints (\ref{constraint}) originates in \cite{Clebsch}.  See \cite{SW68} for an discussion of Clebsch variables and also 
Section 7.1 of the textbook \citet{Mir}.  Finally, the evolution equation for wave amplitude $a_k$ is produced by expressing the cubic terms in the Hamiltonian with solutions to the linear problem represented by the quadratic components of the Hamiltonian.  This is an explicit linearization of the problem prior to the formulation of the kinetic equation.  

\subsubsection{Olbers, McComas and Meiss}
Derivations presented in \cite{O74}, \cite{McC75}, and \cite{Meiss} are based upon the Lagrangian equations of motion:
\begin{eqnarray}
\ddot{x} - f\dot{y} = \frac{-1}{\rho} p_x \nonumber \\
\ddot{y} + f\dot{x} = \frac{-1}{\rho} p_y \nonumber \\
\ddot{z} + g = \frac{-1}{\rho} p_z \nonumber \\
\partial(x_1, x_2, x_3)/\partial(r_1, r_2, r_3) = 1
\end{eqnarray}
expressing momentum conservation and incompressibility.  Here ${\bf r}$ is the initial position of a fluid parcel at ${\bf x}$:  these are Lagrangian coordinates.  In the context of Hamiltonian mechanics, the associated Lagrangian density is:
\begin{eqnarray}
{\cal L} = \frac{1}{2}\rho\left(\dot x_i \dot x_j + \epsilon_{jkl}f_i
\dot x_k x_l \right) - \rho g \delta_{j3}x_j + {\cal P}(J-1)
\nonumber\end{eqnarray} 
where $x_j = x_j({\bf r},t)$ is the instantaneous position of the
parcel of fluid which was initially at ${\bf r}$, ${\cal P}({\bf x})$ is a Lagrange
multiplier corresponding to pressure, and $J = \partial{\bf
  x}/\partial{\bf r}$ is the Jacobian, which ensures the fluid is
incompressible.  

In terms of variables representing a departure from hydrostatic equilibrium:  
$$\xi_j({\bf r},t) = x_j({\bf r},t) - r_j, \ \pi({\bf r},t) = P({\bf x},t) - {\cal P}_k({\bf r}).$$
the Boussinesq Lagrangian density  $\cal L$ for slow variations in background density $\rho$ is:
\begin{equation}
{\cal L} = \frac{1}{2}[ \xi_i^2 +  \epsilon_{jkl}f_i
\dot \xi_k \xi_l  - N^2\xi_3^2 + \pi(\frac{\partial \xi_i}{\partial x_i} + \Delta_{ii} + \Delta )]
\end{equation}
with $\frac{\partial \xi_i}{\partial x_i} + \Delta_{ii} + \Delta$ representing the continuity
equation where $\Delta = det(\partial \xi_i / \partial x_j)$.  

This Lagrangian is then projected onto a single wave amplitude variable $a$ using the linear internal wave constancy relations\footnote{Wave amplitude $a$ is defined so that $a^{\ast}a$ is proportional to wave energy.} based upon plane wave solutions [e.g. \cite{M76}, (2.26)] and a perturbation expansion in wave amplitude is proposed.  This process has two consequences:  The use of internal wave consistency relations places a condition of zero perturbation potential vorticity upon the result, and the expansion places a small amplitude approximation upon the result with ill defined domain of validity relative to the (later) assertion of weak interactions.

The evolution equation for wave amplitude is Lagrange's equation:
\begin{equation}
\frac{d}{dt} \frac{\partial {\cal L}}{\partial \dot{a}_0} - \frac{\partial {\cal L}}{\partial a_0} = 0 
\end{equation}
in which $a_0$ is the zeroth order wave amplitude.  After a series of approximations, this equation is cast into a field variable equation similar to (\ref{fieldequation}). We emphasize that to get there small displacement of parcel of fluid was used, together with the built in assumption of resonant interactions between internal wave modes.  The \citep{LT,LT2} approach is free from such limitations.

\subsubsection{Caillol and Zeitlin}
A non-Hamiltonian kinetic equation for internal waves was derived in \citet{Zeitlin}, their (61) directly from the dynamical equations of motion, without the use of the Hamiltonian structure.   \citet{Zeitlin} invoke the Craya-Herring decomposition for non-rotating flows which enforces a condition of zero perturbation vorticity on the result.  

\subsubsection{Kenyon and Hasselmann}
The first kinetic equations for wave-wave interactions in a continuously stratified ocean appear in
\citet{K66}, \citet{H66} and \citet{K68}.  \citet{K68} states (without detail) that \citet{K66} and \citet{H66} give numerically similar results.  We have found that \citet{K66} differs from the four approaches examined below on one of the resonant manifolds, but have not pursued the question further.  
It is possible this difference results from a typographical error in \citet{K66}.  We have not rederived this non-Hamiltonian representation and thus exclude it from this study.  

\subsubsection{Pelinovsky and Raevsky}
An important  paper on internal waves is \citet{PR77}.  Clebsch variables are used to obtain the
interaction matrix elements for both constant stratification rates, $N=\mathrm{const.}$, 
and arbitrary buoyancy profiles, $N=N(z)$, in a Lagrangian coordinate representation.
Not much details are given, but there are some similarities in appearance with the Eulerian coordinate representation of \citet{Voronovich}. The most significant result is the identification of a scale invariant (non-rotating, hydrostatic) stationary state which we refer to as the Pelinovsky-Raevsky in the companion paper \citep{Theory}.  It is stated in \cite{PR77} that their matrix elements are
equivalent to those derived in their citation [11], which is ~\citet{B75}.
Because \citet{B75} and \cite{PR77} are in Russian and not generally available, we 
refrain from including them in this comparison.  

\subsubsection{Milder}
An alternative Hamiltonian description was developed in
\citet{Milder}, in isopycnal coordinates without assuming a hydrostatic balance.
The resulting Hamiltonian is an iterative expansion in powers of a
small parameter, similar to the case of surface gravity waves. In
principle, that approach may also be used to calculate wave-wave
interaction amplitudes.  Since those calculations were not done
in \citet{Milder}, we do not pursue the comparison further.

\subsection{Weak Turbulence}\label{B2}
Here we derive the kinetic equation following \cite{ZLF}. 
We introduce wave action as
\begin{equation}
n_{\bf p} = \langle a_\p^* a_\p \rangle, \label{WaveAction}\end{equation}
where $\langle\dots\rangle$ means the averaging over statistical ensemble 
of many realizations of the internal waves. 
To derive the time evolution of $n_\p$ we multiply the amplitude equation 
(\ref{fieldequation}) with Hamiltonian (\ref{HAM}) by $a_\p^*$, multiply the amplitude 
evolution equation of $a_\p^{\ast}$ by $a$, subtract the two equations and  average  $\langle\dots\rangle$ the result. We get
\begin{eqnarray}
\frac{\partial n_\p }{\partial t} &=& \Im 
\int  \left( V_{\p_1\p_2}^{\p} J^\p_{\p_1 \p_2} \delta(\p-\p_1-\p_2) \right.  \nonumber\\
   && 
    \left.    -V_{\p \p_1}^{\p_2} J^{\p_2}_{\p \p_1} \delta(\p_2-\p-\p_1)  \right) d\p_1 d\p_2  \nonumber\\
   && 
    \left.    -V_{\p \p_2}^{\p_1} J^{\p_1}_{\p \p_2} \delta(\p_1-\p_2-\p)  \right) d\p_1 d\p_2,
\label{TwoPoints}
\end{eqnarray}
where we introduced a triple correlation function 
\begin{eqnarray} J^{\p}_{\p_1 \p_2}\delta(\p_1-\p-\p_2)\equiv 
\langle a_\p^* a_{\p_1} a_{\p_2} \rangle.\label{TrippleCorrelator} \end{eqnarray}
If we were to have non-interacting fields, i.e. fields with
$V^{\p}_{\p_1 \p_2}$ being zero, this triple correlation function would be
zero.  We then use perturbation expansion in smallness of interactions to calculate
the triple correlation at first order. 
The first order expression for 
$\partial n_\p/\partial t$ therefore requires computing $\partial J^{\p}_{\p_1 \p_2}/\partial t$ to 
first order. To do so we take definition (\ref{TrippleCorrelator}) and 
use (\ref{fieldequation}) with Hamiltonian (\ref{HAM}) and apply 
 $\langle\dots\rangle$ averaging. We get 
\begin{eqnarray}
\left(i\frac{\partial}{\partial t} +
 (\omega_{\p_1}-\omega_{\p_2}-\omega_{\p_3})
\right) J^{\p_1}_{\p_2 \p_3}&& \nonumber\\
&=& \int \left[ -\frac{1}{2} (V^{\p_1}_{\p_4 \p_5})^* J^{\p_4 \p_5}_{\p_2 \p_3}\delta(\p_1-\p_4-\p_5)  
\right. \nonumber\\ &&\left.
              +(V^{\p_4}_{\p_2 \p_5})^* J^{\p_1 \p_5}_{\p_3 \p_4}\delta(\p_4-\p_2-\p_5) 
\right. \nonumber\\ &&\left.
+V^{\p_4}_{\p_3 \p_5} J^{\p_1 \p_5}_{\p_2 \p_4}\delta(\p_4-\p_3-\p_5) \right] d\p_4 d\p_5.
\label{TrippleCorrelatorddt}
\end{eqnarray} 
Here we introduced the quadruple correlation function             
\begin{eqnarray} J^{\p_1 \p_2}_{\p_3 \p_4}\delta(\p_1+\p_2-\p_3-\p_4)\equiv 
\langle a_{\p_1}^*  a_{\p_2}^*   a_{\p_3} a_{\p_4} \rangle.
\label{FourCorrelator} \end{eqnarray}
The next step is to assume Gaussian statistics, and to express $J^{\p_1 \p_2}_{\p_3 \p_4}$
as a product of two two-point correlators as
$$J^{\p_1 \p_2}_{\p_3 \p_4} = n_{\p_1} n_{\p_2} \Big[
\delta(\p_1-\p_3)\delta(\p_2-\p_4) + 
\delta(\p_1-\p_4)\delta(\p_2-\p_3)\Big].$$
Then 
\begin{eqnarray}
\left[
i\frac{\partial }{\partial t} +  (\omega_{\p_1}-\omega_{\p_2}-\omega_{\p_3})
\right] J^{\p_1}_{\p_2\p_3} 
= (V^{\p_1}_{\p_2 \p_3})^* \left(n_1 n_3 + n_1 n_2 - n_2 n_3 \right).\label{TripplePointGauss}
\end{eqnarray}
Time integration of the equation for $J^{\p_1}_{\p_2\p_3}$ will contain fast
oscillations due to initial value of $J^{\p_1}_{\p_2\p_3}$ and slow
evolution due to the nonlinear wave interactions. Contribution from 
first term will rapidly decrease with time, so neglecting these terms we 
get 
\begin{eqnarray}
J^{\p_1}_{\p_2\p_3} 
= \frac{
(V^{\p_1}_{\p_2 \p_3})^* \left(n_1 n_3 + n_1 n_2 - n_2 n_3 \right)}
{\omega_{\p_1}-\omega_{\p_2}-\omega_{\p_3} + i \Gamma_{\p_1\p_2\p_3}}.
\label{TheMeat}
\end{eqnarray}
Here we introduced the nonlinear damping of the waves $\Gamma_{\p_1\p_2\p_3}$. 
We will elaborate on $\Gamma_{\p_1\p_2\p_3}$ in Section (\ref{OffResonant}). 
We now  substitute (\ref{TheMeat}) into (\ref{TwoPoints}), 
assume for now that the damping of the wave is small, and use
\begin{equation}
\lim_{\Delta\to 0} \Im \left[\frac{1}{\Delta+i\Gamma} \right]= - \pi \delta(\Delta).
\label{DeltaFunction}
\end{equation}
We then obtain the three-wave kinetic equation~\citep{ZLF,NoisyNazarenko,LLNZ}:
\begin{eqnarray}
\frac{d n_{{\bf p}}}{dt} = 4\pi \int
 |V_{{\bf p}_1,{\bf p}_2}^{{\bf p}}|^2 \, f_{p12} \,
\delta_{{{\bf p} - {\bf p}_1-{\bf p}_2}} \, \delta({\omega_{{\bf p}}
-\omega_{{{\bf p}_1}}-\omega_{{{\bf p}_2}}})
d {\bf p}_{12}
\nonumber \\
-4\pi\int
 \, |V_{{\bf p}_2,{\bf p}}^{{\bf p}_1}|^2\, f_{12p}\, \delta_{{{\bf p}_1 - {\bf p}_2-{\bf p}}} \,
  \delta({{\omega_{{\bf p}_1} -\omega_{{\bf p}_2}-\omega_{{\bf p}}}})
\, d {\bf p}_{12}
\nonumber \\
-4\pi\int
 \, |V_{{\bf p},{\bf p}_1}^{{\bf p}_2}|^2\, f_{2p1}\, \delta_{{{\bf p}_2 - {\bf p}-{\bf p}_1}} \,
  \delta({{\omega_{{\bf p}_2} -\omega_{{\bf p}}-\omega_{{\bf p}_1}}})
\, d {\bf p}_{12}
 \, ,\nonumber\\
{\rm with} ~~ f_{p12} = n_{{\bf p}_1}n_{{\bf p}_2} -
n_{{\bf p}}(n_{{\bf p}_1}+n_{{\bf p}_2}) \, .
\label{KineticEquation}
\end{eqnarray}

Here $n_{{\bf p}} = n({\bf p})$ is a three-dimensional wave action spectrum (spectral energy density divided by frequency) and the interacting wavevectors ${\bf p}$, ${\bf p}_1$ and ${\bf p}_2$ are given by
$${\bf p} = ({\bf k}, m),$$
i.e.\ ${\bf k}$ is the horizontal part of ${\bf p}$ and $m$ is its vertical component.  We assume the wavevectors are signed variables and wave frequencies $\omega_{{\bf p}}$ are restricted to be positive.  The magnitude of wave-wave interactions 
$V_{{\bf p},{\bf p}_1}^{{\bf p}_2}$
is a matrix representation of the coupling between triad members.  
 It serves as a multiplier in the nonlinear convolution term in what is now commonly called the Zakharov equation -- equation in the Fourier space for the waves field variable. This is also an expression that multiplies the cubic convolution term in the three-wave Hamiltonian. 


We re-iterate that typical assumptions needed for the derivation of
kinetic equations are:
\begin{itemize}
\item Weak nonlinearity, 
\item Gaussian statistics of the interacting wave field in wavenumber space and
\item Resonant wave-wave interactions
\end{itemize}
We note that the derivation given here is schematic.  A more systematic derivation can be obtained using only an assumption of weak nonlinearity.

\subsection{The Boltzmann Rate}
The kinetic equation allows us to numerically estimate the life time
of any given spectrum. In particular, we can define a wavenumber dependent nonlinear time scale proportional to the inverse Boltzmann rate:
\begin{equation}
\tau^{\mathrm{NL}}_{{\bf p}} = \frac{n_{{\bf p}}}{\dot n_{{\bf p}}}~.
\label{NonlinearTime}
\end{equation}
This time scale characterizes the net rate at which the spectrum changes and can 
be directly calculated from the kinetic equation.  

One can also define the 
characteristic linear time scale, equal to a wave period  
$$\tau^{\mathrm{L}}_{{\bf p}}=2\pi/\omega_{{\bf p}}.$$ 
The non-dimensional ratio of these time scales can 
characterize  the level of nonlinearity in the nonlinear system:
\begin{equation}
{\cal  \epsilon}_{{\bf p}} = 
\frac
{\tau_{{\bf p}}^{\mathrm{L}}} 
{\tau_{{\bf p}}^{\mathrm{NL}}}
= 
\frac
{2\pi \dot n_{{\bf p}}}
{n_{{\bf p}} \omega_{{\bf p}}}
\label{NonlinearRatio}
\end{equation}
We refer to (\ref{NonlinearRatio}) as a normalized Boltzmann rate. 

The normalized Boltzmann rate serves as a low order 
consistency check for the various kinetic equation derivations. An $O(1)$ value of ${\cal  \epsilon}_{{\bf p}}$ implies that the derivation of the kinetic equation is internally inconsistent.  The Boltzmann rate represents the {\em net} rate of transfer for wavenumber ${\bf p}$.  The individual rates of transfer into and out of ${\bf p}$ (called Langevin rates) are typically greater than the Boltzmann rate, \citep{M86, PMW80}.  This is particularly true in the Induced Diffusion regime (defined below in Section \ref{Special}) in which the rates of transfer into and out of ${\bf p}$ are one to three orders of magnitude larger than their residual and the Boltzmann rates we calculate are not appropriate for either spectral spike or potentially for smooth, homogeneous but anisotropic spectra \citep{M86}.  Estimates of the individual rates of transfer into and out of ${\bf p}$ can be addressed through Langevin methods \citep{PMW80}.  We focus here simply on the Boltzmann rate to demonstrate inconsistencies with the assumption of a slow time evolution.  Estimates of the Boltzmann rate and ${\cal  \epsilon}_{{\bf p}}$  require integration of (\ref{KineticEquation}).  In this manuscript such integration is performed numerically.

\section{Resonant wave-wave interactions - nonrotational limit}\label{ResonantInteractions} 

How one can compare the function of two vectors ${\bf p}_1$ and ${\bf p}_2$, and their
sum or difference? First one realizes that out of 6 components of ${\bf p}_1$ and ${\bf p}_2$, only relative angles between wavevectors enter into the equation for matrix elements. That is because the matrix elements depend on the inner and outer products of wavevectors. The overall horizontal orientation of the wavevectors does not matter:  relative angles can be determined from a triangle inequality and the magnitudes of the horizontal wavevectors ${\bf k}$, ${\bf k}_1$ and ${\bf k}_2$. Thus the only needed components are $|{\bf k}|$, $|{\bf k}_1|$, $|{\bf k}_2|$, $m$ and $m_1$ ($m_2$ is computed from $m$ and $m_1$).  Further note that in the $f=0$ and hydrostatic limit, all matrix elements become scale invariant functions.  It is therefore sufficient to choose an arbitrary scalar value for $|{\bf k}|$, and $m$, since only $|{\bf k}_1|/|{\bf k}|$, $|{\bf k}_2|/|{\bf k}|$ and  $m_1/m$ enter the expressions for matrix elements.  We make the particular (arbitrary) choice that $|{\bf k}|=m=1$ for the purpose of numerical evaluation, and thus the only independent variables to consider are $|{\bf k}_1|$, $|{\bf k}_2|$ and $m_1$. Finally, $m_1$ is determined from the resonance conditions, as explained in the next subsection below. As a result, we are left with a matrix element as a function of only two parameters, $k_1$ and $k_2$.  This allows us to easily compare the values of matrix elements on the resonant manifold by plotting the values as a function of the two parameters.  

\subsection{Reduction to the Resonant Manifold}
When confined to the traditional form of the kinetic equation, wave-wave
interactions are constrained to the resonant manifolds defined by
\begin{eqnarray}
a)~~
\begin{cases}
 \p = \p_1 + \p_2 \\
 \omega = \omega_1 + \omega_2
\end{cases}
b)~~
\begin{cases}
 \p_1 = \p_2 + \p \\
 \omega_1 = \omega_2 + \omega
\end{cases}
c)~~
\begin{cases}
 \p_2 = \p + \p_1 \\
 \omega_2 = \omega + \omega_1
\end{cases}.
\label{RESONANCES}
\end{eqnarray}
To compare matrix elements on the resonant manifold we are going to
use the above resonant conditions and the internal-wave dispersion relation (\ref{dispersion}).
To determine vertical components $m_1$ and $m_2$ of the interacting
wavevectors, one has to solve the resulting quadratic equations. Without restricting generality we choose  $m>0$.  There are two solutions for $m_1$ and $m_2$ given below for each of the three resonance types described above.

Resonances of type (\ref{RESONANCES}a) give
\begin{subequations}
	 \allowdisplaybreaks
 \begin{align}
&
\begin{cases}
 m_1 = \frac{m}{2 |{\bf k}|} \left(|{\bf k}| + |{\bf k}_1| + |{\bf k}_2| + \sqrt{(|{\bf k}| + |{\bf k}_1| + |{\bf k}_2|)^2 - 4 |{\bf k}| |{\bf k}_1|}\right)
 \\
 m_2 = m - m_1.
\end{cases}
,
\label{eq:sol1}
\\
&
\begin{cases}
m_1 = \frac{m}{2|{\bf k}|} \left(|{\bf k}| - |{\bf k}_1| - |{\bf k}_2| - \sqrt{(|{\bf k}| - |{\bf k}_1| - |{\bf k}_2|)^2 + 4 |{\bf k}| |{\bf k}_1|}\right)
\\
m_2 = m - m_1.
\end{cases}
,
\label{eq:sol2}
 \end{align}
	\end{subequations}
Note that because of the symmetry, (\ref{eq:sol1}) translates to (\ref{eq:sol2}) if wavenumbers $1$ and $2$ are exchanged.
 
Resonances of type (\ref{RESONANCES}b) give
\begin{subequations}
\allowdisplaybreaks
\begin{align}
&
\begin{cases}
m_2 = - \frac{m}{2 |{\bf k}|} \left(|{\bf k}| - |{\bf k}_1| - |{\bf k}_2| + \sqrt{(|{\bf k}| - |{\bf k}_1| - |{\bf k}_2|)^2 + 4 |{\bf k}| |{\bf k}_2|}\right)
\\
m_1 = m + m_2.
\end{cases}
,
\label{eq:sol3}
\\
&
\begin{cases}
m_2 = - \frac{m}{2|{\bf k}|} \left(|{\bf k}| + |{\bf k}_1| - |{\bf k}_2| + \sqrt{(|{\bf k}| + |{\bf k}_1| - |{\bf k}_2|)^2 + 4 |{\bf k}| |{\bf k}_2|}\right)
\\
m_1 = m + m_2.
\end{cases}
,
\label{eq:sol4}
\end{align}
\end{subequations}

Resonances of type (\ref{RESONANCES}c) give
\begin{subequations}
\allowdisplaybreaks
\begin{align}
&
\begin{cases}
m_1 = - \frac{m}{2|{\bf k}|} \left(|{\bf k}| - |{\bf k}_1| - |{\bf k}_2| + \sqrt{(|{\bf k}| - |{\bf k}_1| - |{\bf k}_2|)^2 + 4 |{\bf k}| |{\bf k}_1|}\right)
\\
m_2 = m + m_1.
\end{cases}
,
\label{eq:sol5}
\\
&
\begin{cases}
m_1 = - \frac{m}{2|{\bf k}|} \left(|{\bf k}| - |{\bf k}_1| + |{\bf k}_2| + \sqrt{(|{\bf k}| - |{\bf k}_1| + |{\bf k}_2|)^2 + 4 |{\bf k}| |{\bf k}_1|}\right)
\\
m_2 = m + m_1.
\end{cases}
.
\label{eq:sol6}
\end{align}
\end{subequations}
Because of the symmetries of the problem, (\ref{eq:sol3}) is equivalent to
(\ref{eq:sol5}), and (\ref{eq:sol4}) is equivalent to (\ref{eq:sol6})
if wavenumbers $1$ and $2$ are exchanged.

\subsection{Comparison of matrix elements}

As explained above, we assume $f=0$ and hydrostatic balance. Such a
choice makes the matrix elements to be scale-invariant functions
that depend only upon $|{\bf k}_1|$ and $|{\bf k}_2|$.
As a consequence of the triangle inequality we need to consider matrix elements only within a ``kinematic box'' defined by $$||{\bf k}_1| - |{\bf k}_2|| < |{\bf k}| < |{\bf k}_1| + |{\bf k}_2|.$$
The matrix elements will have different values depending on the dimensions so that isopycnal and Eulerian approaches will give different values (\ref{eq:dispersionISO})-(\ref{eq:dispersionCAR}). To address this issue in the simplest possible way, we multiply each matrix element by a dimensional number chosen so that all
 matrix elements are equivalent for some specific wavevector. In particular, we choose the scaling constant so 
that $|V(|{\bf k}_1|=1,|{\bf k}_2|=1)|^2=1$. This allows a transparent comparison without worrying about dimensional 
differences between various formulations.

\subsubsection{Resonances of the ``sum'' type (\ref{RESONANCES}a)}

\begin{figure*}[t]
  \noindent\includegraphics[width=19pc,angle=0]{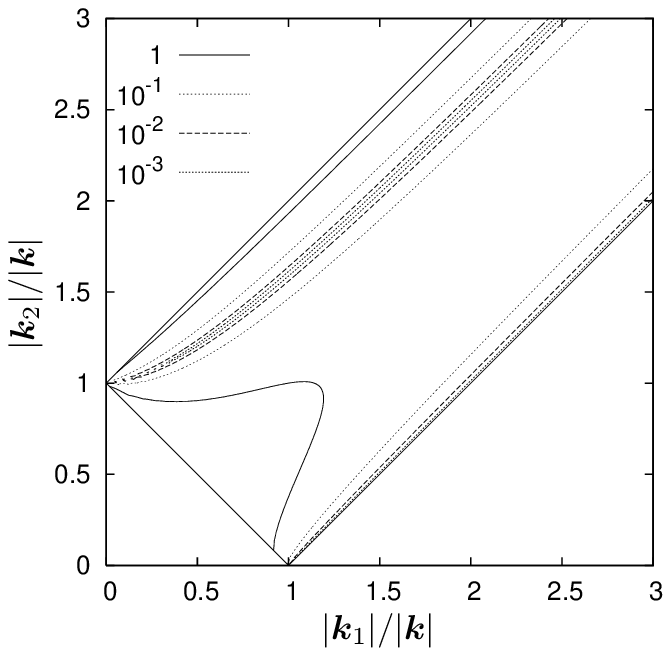} 
  \noindent\includegraphics[width=19pc,angle=0]{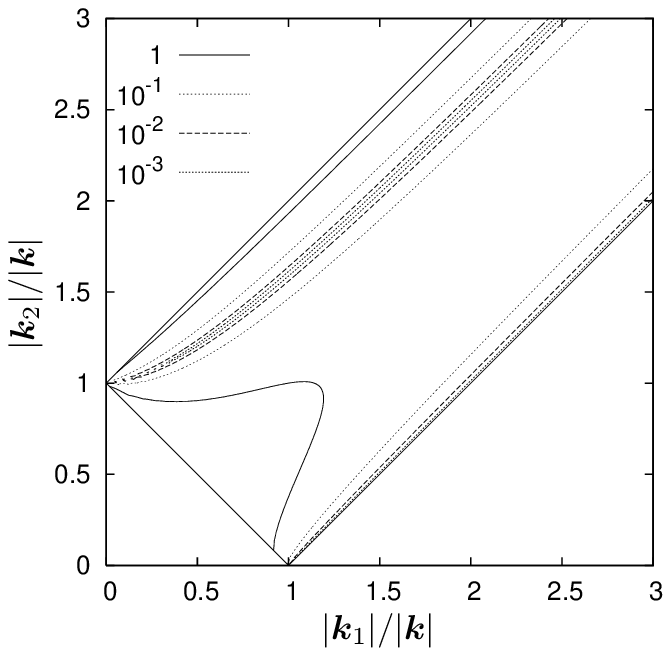} \\  
  \noindent\includegraphics[width=19pc,angle=0]{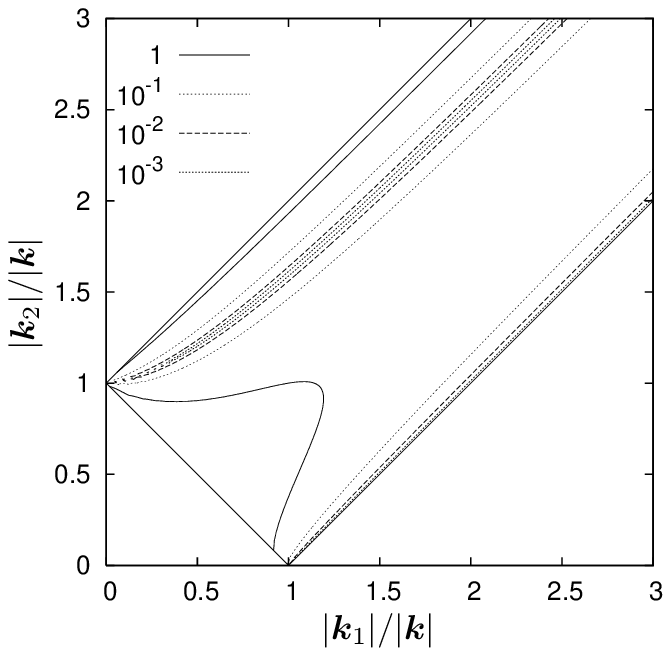} 
  \noindent\includegraphics[width=19pc,angle=0]{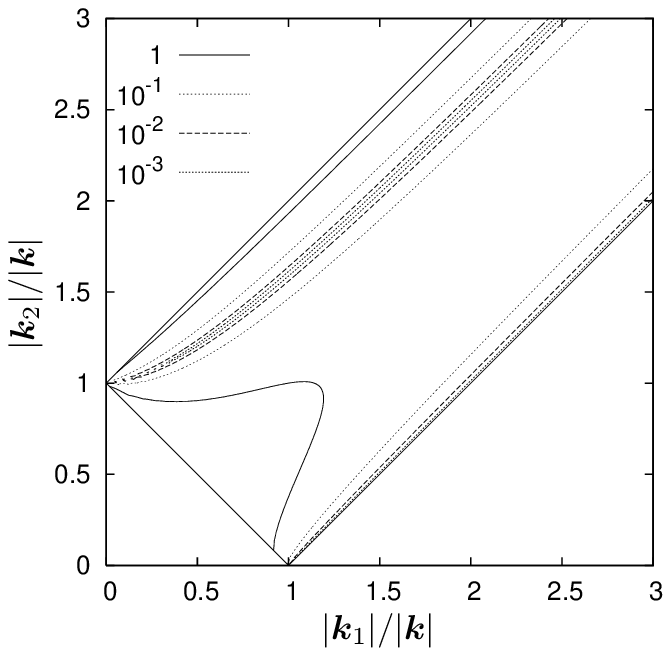} \\
\caption{Matrix elements   $\mid{V^{\p}_{\p_1, \p_2}}_ {\mathrm{(\ref{eq:sol2})}} \mid^2$ given by the solution (\ref{eq:sol2}).
upper left: $\mid{V^{\p}_{\p_1, \p_2}}_{\mathrm{(\ref{eq:sol2})}}^{{\rm MO}}\mid^2$ according to \citet{MO75},
upper right: $\mid{V^{\p}_{\p_1, \p_2}}_{\mathrm{(\ref{eq:sol2})}}^{{\rm V}}\mid^2$ according to \citet{Voronovich},
bottom left: $\mid{V^{\p}_{\p_1, \p_2}}_{\mathrm{(\ref{eq:sol2})}}^{{\rm CZ}}\mid^2$ according to \citet{Zeitlin},
bottom right: $\mid{V^{\p}_{\p_1, \p_2}}_{\mathrm{(\ref{eq:sol2})}}^{{\rm H}}\mid^2$ according to \citet{LT}.
}
\label{FIGRESONANTa}
\end{figure*}

Figure~\ref{FIGRESONANTa}
presents the values of the matrix element 
$|{V^{{\bf p}}_{{\bf p}_1, {\bf p}_2}}_{\mathrm{(\ref{eq:sol2})}}|^2$ on the resonant sub-manifold given explicitly by
(\ref{eq:sol2}).
All approaches
give equivalent
results. This is confirmed by plotting the relative ratio between
these approaches, and it is given by numerical noise (not shown).
The solution (\ref{eq:sol1}) gives the same matrix elements
but with $|{\bf k}_1|$ and $|{\bf k}_2|$ exchanged
owing to their symmetries.

\subsubsection{Resonances of the ``difference'' type (\ref{RESONANCES}b) and
 (\ref{RESONANCES}c)}

\begin{figure*}[t]
  \noindent\includegraphics[width=19pc,angle=0]{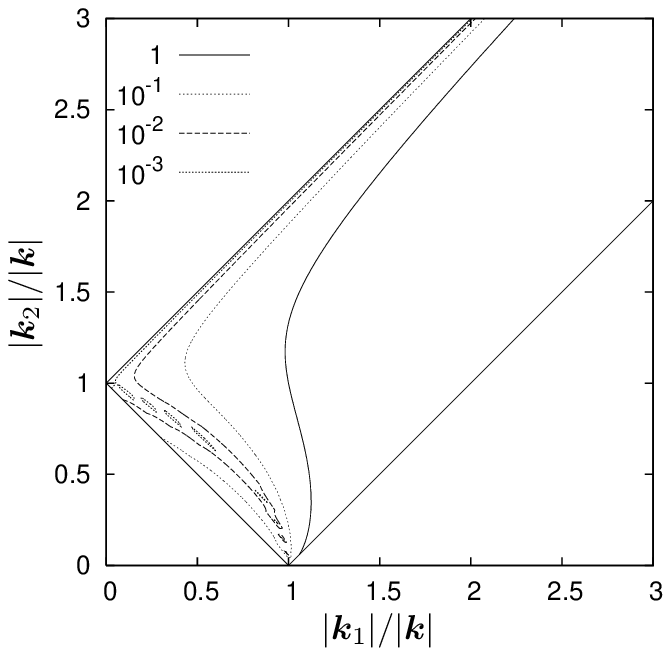} 
  \noindent\includegraphics[width=19pc,angle=0]{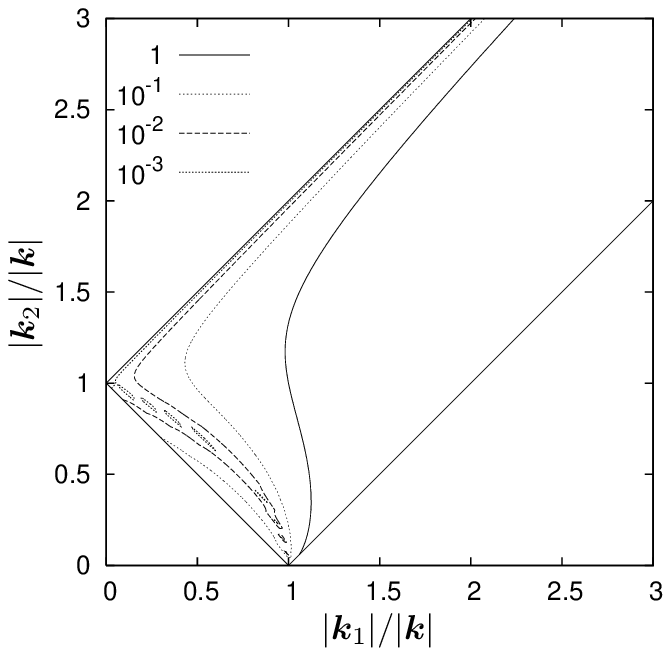} \\
  \noindent\includegraphics[width=19pc,angle=0]{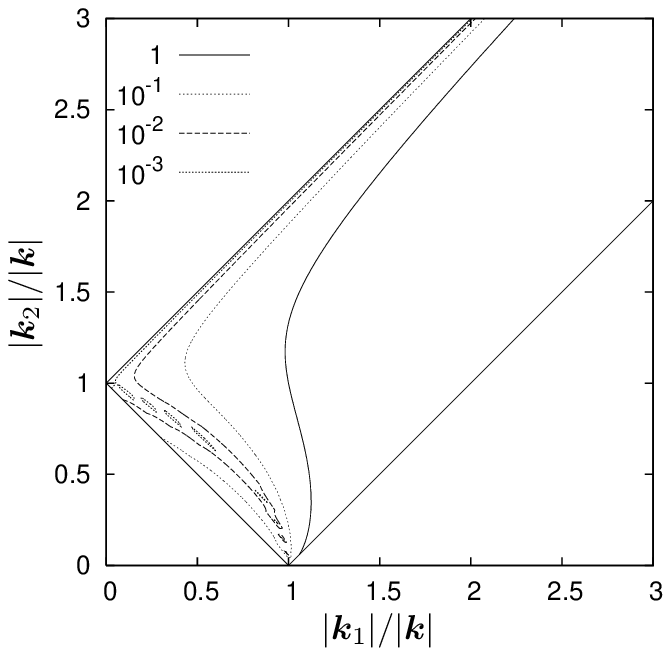} 
  \noindent\includegraphics[width=19pc,angle=0]{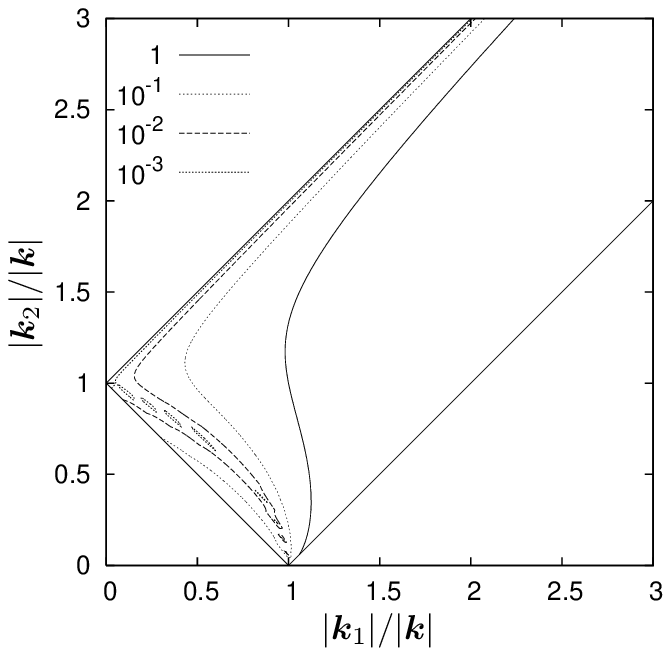} \\
\caption{Matrix elements $\mid{V^{\p_1}_{\p_2, \p}}_{\mathrm{(\ref{eq:sol3})}}\mid^2$ given by the solution (\ref{eq:sol3}).
upper left: $\mid{V^{\p_1}_{\p_2, \p}}_{\mathrm{(\ref{eq:sol3})}}^{\mathrm{MO}}\mid^2$ according to \citet{MO75},
upper right: $\mid{V^{\p_1}_{\p_2, \p}}_{\mathrm{(\ref{eq:sol3})}}^{\mathrm{V}}\mid^2$ according to \citet{Voronovich},
bottom left: $\mid{V^{\p_1}_{\p_2, \p}}_{\mathrm{(\ref{eq:sol3})}}^{\mathrm{CZ}}\mid^2$ according to \citet{Zeitlin},
bottom right: $\mid{V^{\p_1}_{\p_2, \p}}_{\mathrm{(\ref{eq:sol3})}}^{\mathrm{H}}\mid^2$ according to \citet{LT}.
}
\label{FIGRESONANTb}
\end{figure*}

\begin{figure*}[t]
  \noindent\includegraphics[width=19pc,angle=0]{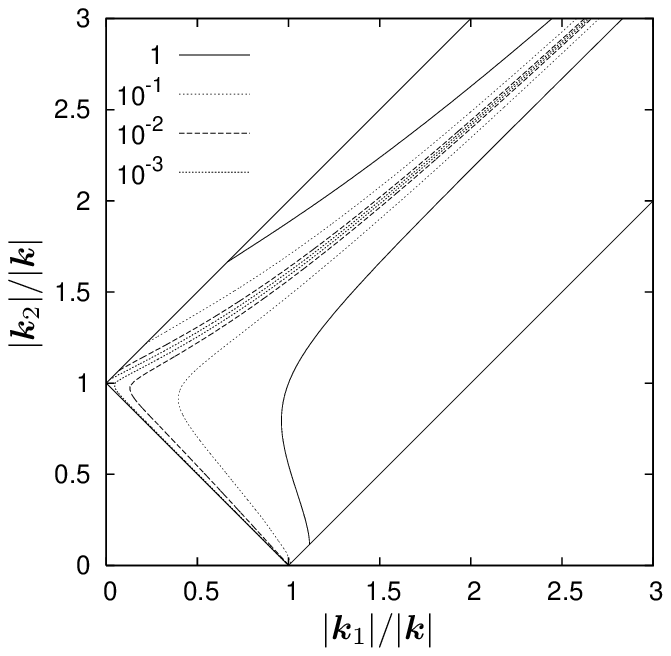}
  \noindent\includegraphics[width=19pc,angle=0]{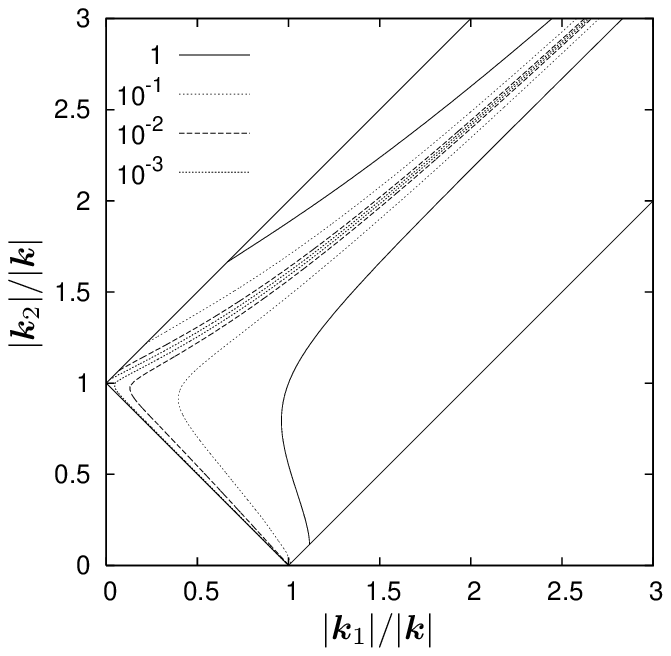} \\
  \noindent\includegraphics[width=19pc,angle=0]{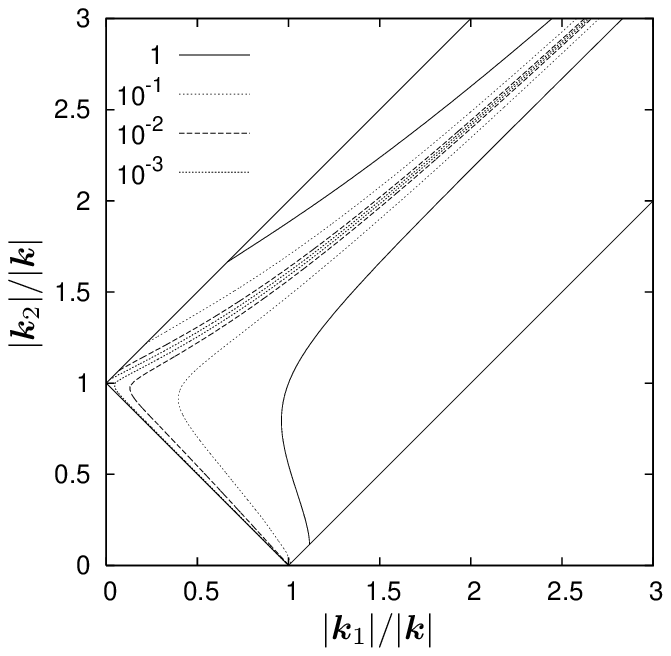} 
  \noindent\includegraphics[width=19pc,angle=0]{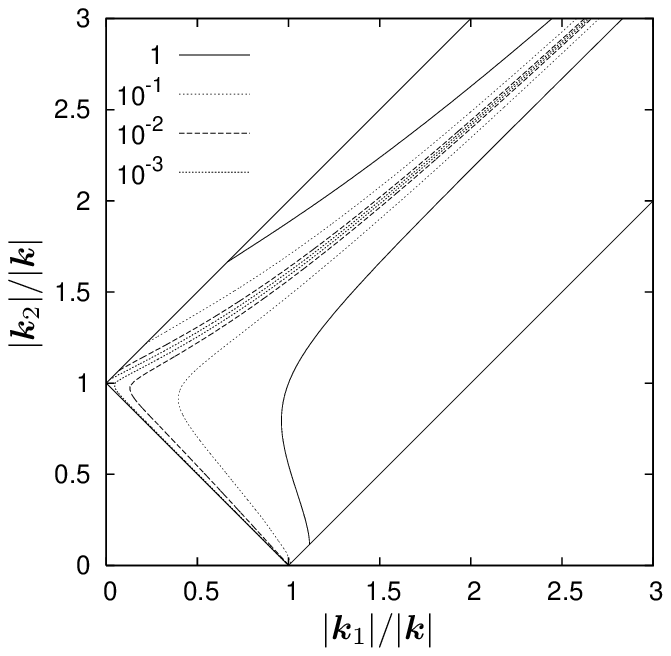} \\
\caption{Matrix elements 
$\mid{V^{\p_1}_{\p_2, \p}}_{\mathrm{(\ref{eq:sol4})}}|^2$ given by the solution (\ref{eq:sol4}).
upper left: $\mid{V^{\p_1}_{\p_2, \p}}_{\mathrm{(\ref{eq:sol4})}}^{\mathrm{MO}}\mid^2$ according to \citet{MO75},
upper right: $\mid{V^{\p_1}_{\p_2, \p}}_{\mathrm{(\ref{eq:sol4})}}^{\mathrm{V}}\mid^2$ according to \citet{Voronovich},
bottom left: $\mid{V^{\p_1}_{\p_2, \p}}_{\mathrm{(\ref{eq:sol4})}}^{\mathrm{CZ}}\mid^2$ according to \citet{Zeitlin},
bottom right: $\mid{V^{\p_1}_{\p_2, \p}}_{\mathrm{(\ref{eq:sol4})}}^{\mathrm{H}}\mid^2$ according to \citet{LT}.
}
\label{FIGRESONANTc}
\end{figure*}

We then turn our attention to resonances of ``difference'' type (\ref{RESONANCES}b) for which 
(\ref{RESONANCES}c) could be obtained by symmetrical exchange of the indices.
All the matrix elements 
$|{V^{{\bf p}_1}_{{\bf p}_2, {\bf p}}}_{\mathrm{(\ref{eq:sol3})}}|^2$
on the resonant sub-manifold (\ref{eq:sol3}), 
are shown in Fig.~\ref{FIGRESONANTb}.
All the matrix elements are equivalent.  The relative differences between different approaches are
given by numerical noise (not shown).
Finally, $|{V^{{\bf p}_1}_{{\bf p}_2, {\bf p}}}_{\mathrm{(\ref{eq:sol4})}}|^2$ on the 
resonant sub-manifold (\ref{eq:sol4}) are shown in Fig.~\ref{FIGRESONANTc}.
Again, all the matrix elements
are equivalent.

The solutions (\ref{eq:sol5}) and (\ref{eq:sol6}) give the same matrix elements but with $|{\bf k}_1|$ and $|{\bf k}_2|$ exchanged
as the solutions (\ref{eq:sol3}) and (\ref{eq:sol4})
owing to their symmetries.

\subsubsection{Special triads}\label{Special}
Three simple interaction mechanisms are identified by \citet{MB77} in the limit of an extreme scale separation.  In this subsection we look in closer detail at these special limiting triads to confirm that all matrix elements are indeed asymptotically consistent.  The limiting cases are:

\begin{itemize}
\item
the vertical backscattering of a high-frequency wave by a low frequency wave of twice the vertical wavenumber into a second high-frequency wave of oppositely signed vertical wavenumber and nearly the same wavenumber magnitude. This type of scattering is called elastic scattering (ES). The solution (\ref{eq:sol1}) in the limit  $|{\bf k}_1| \to 0$ corresponds to this type of special triad.
\item
The scattering of a high-frequency wave by a low-frequency, small-wavenumber wave into a second, nearly identical, high-frequency large-wavenumber wave. This type of scattering is called  induced diffusion (ID). The solution (\ref{eq:sol2}) in the limit that $|{\bf k}_1| \to 0$ corresponds to this type of special triad.
\item
The decay of a low wavenumber wave into two high vertical wavenumber waves of approximately one-half the frequency. This is called  parametric subharmonic instability (PSI).  The solution (\ref{eq:sol3}) in the limit that $|{\bf k}_1| \to 0$ corresponds to this type of triad.
\end{itemize}

\begin{figure*}[tp]
 \begin{center}
  \noindent\includegraphics[width=19pc,angle=0]{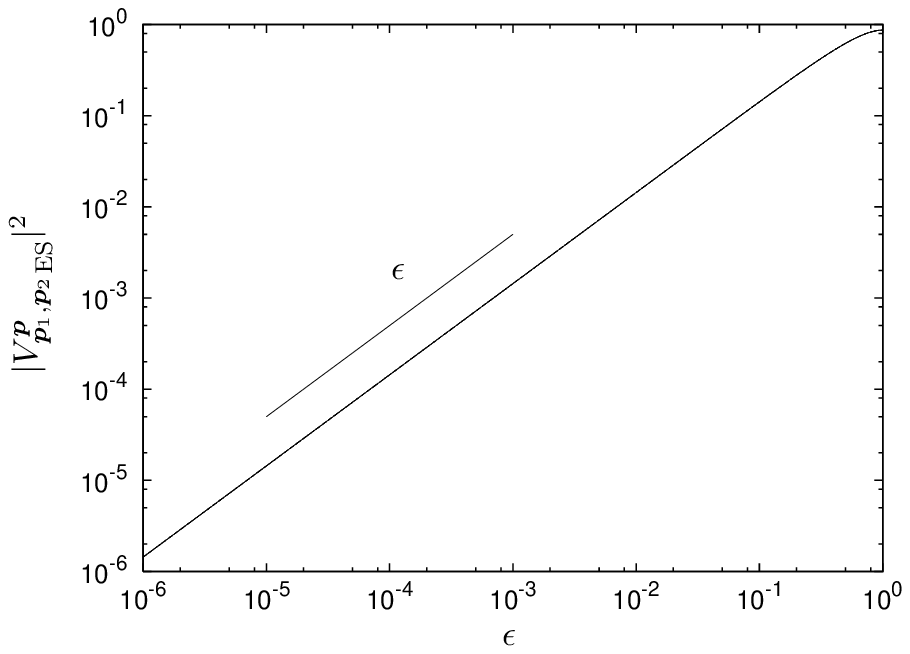} \\
  \noindent\includegraphics[width=19pc,angle=0]{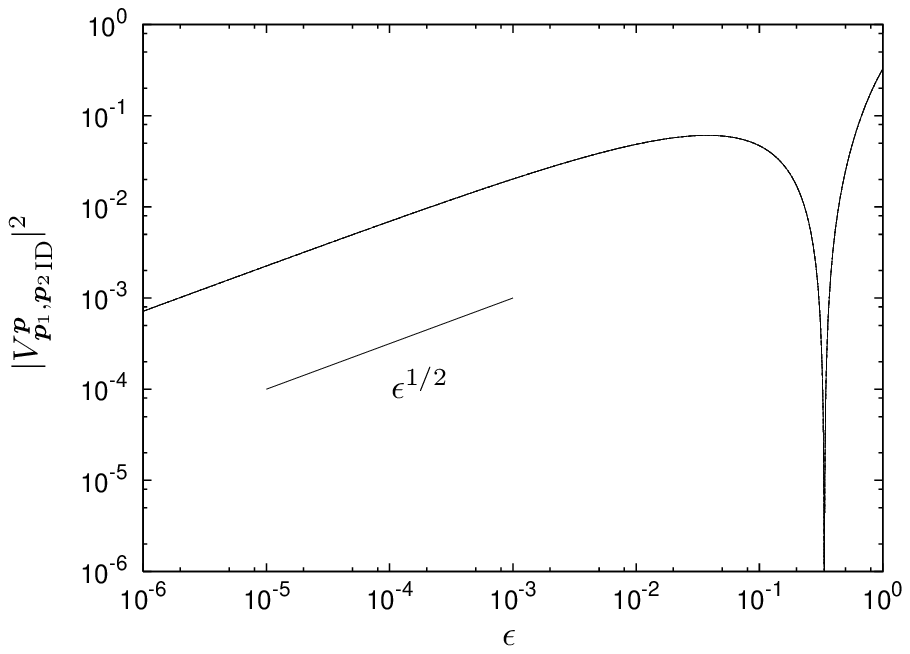} \\
  \noindent\includegraphics[width=19pc,angle=0]{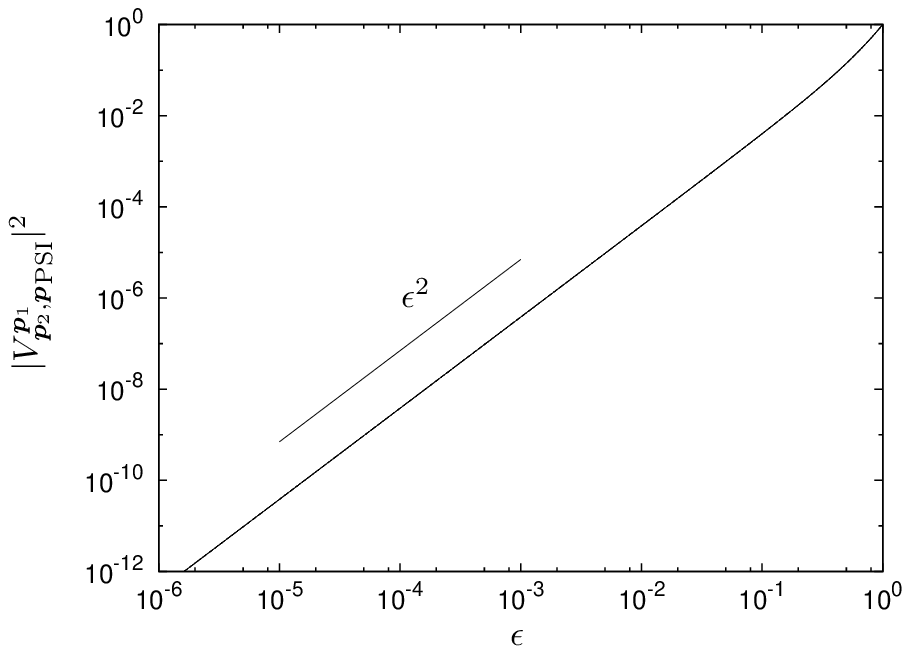} \\
  \caption{upper: Matrix elements $\mid{V^{\p}_{\p_1, \p_2}}_{\mathrm{ES}}\mid^2$ given by the solution (\ref{eq:sol1}).
middle: Matrix elements $\mid{V^{\p}_{\p_1, \p_2}}_{\mathrm{ID}}\mid^2$ given by the solution (\ref{eq:sol2}).
  bottom: Matrix elements $\mid{V^{\p_1}_{\p_2, \p}}_{\mathrm{PSI}}\mid^2$ given by the solution (\ref{eq:sol3}),
  which gives PSI as $\mid{\bf k}_1\mid \to 0$ ($\epsilon \to 0$).
The matrix elements here are shown as functions of $\epsilon$ such that $(|{\bf k}_1|, |{\bf k}_2|) = (\epsilon, \epsilon/3+1) |{\bf k}|$. All four versions of the Matrix elements are plotted here:  the appearance of a single line in each figure panel testifies to the similarity of the elements on the resonant manifold.  
}
\label{FigureThree}
 \end{center}
\end{figure*}

To study the detailed behavior of the matrix elements in the special triad cases, we choose to present the matrix elements along a straight line defined by
$$
(|{\bf k}_1|, |{\bf k}_2|) = (\epsilon, \epsilon/3+1)  |{\bf k}|.
$$
This line is defined in such a way so that it originates from the corner of the kinematic box in Figs.~\ref{FIGRESONANTa}--\ref{FIGRESONANTc} at $(|{\bf k}_1|, |{\bf k}_2|) = (0,|{\bf k}|) $ and has a slope of 1/3.  The slope of this line is arbitrary.  We could have taken $\epsilon/4$ or $\epsilon/2$.  The matrix elements here are shown as functions of $\epsilon$
in Fig.~\ref{FigureThree}.  We see that
all four approaches are again {\em equivalent}  on the resonant manifold for the 
case of special triads.

In this section we demonstrated that all four approaches we
considered produce {\it equivalent} results on the resonant
manifold in the absence of background rotation. This statement is not 
trivial, given the different assumptions and coordinate systems that
have been used for the various kinetic equation derivations.

\section{Resonant wave-wave interactions \label{Rotations} - 
in the presence of Background Rotations}

In the presence of background rotation, the matrix elements loose their scale invariance due to the introduction of an additional time scale ($1/f$) in the system.  Consequently the comparison of matrix elements is performed as a function of four independent parameters.  

We perform this comparison in the frequency-vertical wavenumber domain.  In particular, for arbitrary $\omega$, $\omega_1$, $m$ and $m_1$, $\omega_2$ and $m_2$ can be calculated by requiring that they satisfy the resonant conditions $\omega = \omega_1 + \omega_2$ and $m = m_1 + m_2$.  We then can check whether the corresponding horizontal wavenumber magnitudes $k$, given by
\begin{eqnarray}
k_i & = &  \frac{m_i N \rho_o }{g} \sqrt{\omega_i^2 - f^2} {\rm ~~(isopycnal ~coordinates)~~and~~ } \nonumber \\ 
k_i & = &  m_i\frac{\sqrt{\omega_i^2 - f^2}}{N} {\rm ~~(Lagrangian ~ coordinates)} 
\end{eqnarray} 
satisfy the triangle inequality.  The matrix elements of the isopycnal 
and Lagrangian 
coordinate representations are then calculated.  We are performed this comparison for $10^{12}$ points on the resonant manifold.  After being multiplied by an appropriate dimensional number to convert between Eulerian and isopycnal coordinate systems, the two matrix elements coincide up to machine precision. 

One might, with sufficient experience, regard this as an intuitive statement.  It is, however, far from trivial given the different assumptions and coordinate representations.  In particular, we note that derivations of the wave amplitude evolution equation in Lagrangian coordinates \citep{O76, McC75, Meiss} do not explicitly contain a potential vorticity conservation statement corresponding to assumption (\ref{PV}) in the isopycnal coordinate \citep{LT2} derivation.  We have inferred that the Lagrangian coordinate derivation conserves potential vorticity as that system is projected upon the linear modes of the system having zero perturbation potential vorticity.

\section{Resonance Broadening and Numerical Methods}\label{RBnum}

\subsection{Nonlinear frequency renormalization as a result of nonlinear wave-wave interactions}\label{OffResonant}

The resonant interaction approximation is a self-consistent mathematical simplification which reduces the complexity of the problem for weakly nonlinear systems.  As nonlinearity increases, near-resonant interactions become more and more pronounced and need to be addressed.  Moreover, near-resonant
interactions play a major role in numerical simulations on a discrete
grids~\citep{LvovNazarenkoPokorni}, for time evolution of discrete systems
~\citep{Gersh2007}, in acoustic turbulence ~\citep{LLNZ}, surface
gravity waves~\citep{JansenXXX,yuen-lake}, and internal waves
\citep{Voron2006,Shrira}.

To take into account the effects of near-resonant interactions
self-consistently, we revisit Section \ref{B2}. Now we {\it do
  not } take the limit $\Gamma_{\p\p_1\p_2}\to 0$. Then, instead of the kinetic
equation with the frequency conserving delta-function, we obtain the 
{\it generalized} kinetic equation
\begin{eqnarray}
\frac{d n_{{\bf p}}}{dt} = 4 \int
 |V_{{\bf p}_1,{\bf p}_2}^{{\bf p}}|^2 \, f_{p12} \,
\delta_{{{\bf p} - {\bf p}_1-{\bf p}_2}} \, {\cal L}({\omega_{{\bf p}}
-\omega_{{{\bf p}_1}}-\omega_{{{\bf p}_2}}})
d {\bf p}_{12}
\nonumber \\
-4\int
 \, |V_{{\bf p}_2,{\bf p}}^{{\bf p}_1}|^2\, f_{12p}\, \delta_{{{\bf p}_1 - {\bf p}_2-{\bf p}}} \,
  {\cal L} ({{\omega_{{\bf p}_1} -\omega_{{\bf p}_2}-\omega_{{\bf p}}}})
\, d {\bf p}_{12}
\nonumber \\
-4\int
 \, |V_{{\bf p},{\bf p}_1}^{{\bf p}_2}|^2\, f_{2p1}\, \delta_{{{\bf p}_2 - {\bf p}-{\bf p}_1}} \,
  {\cal L}({{\omega_{{\bf p}_2} -\omega_{{\bf p}}-\omega_{{\bf p}_1}}})
\, d {\bf p}_{12}
 \, ,\nonumber\\
{\rm with} ~~ f_{p12} = n_{{\bf p}_1}n_{{\bf p}_2} -
n_{{\bf p}}(n_{{\bf p}_1}+n_{{\bf p}_2}) \, ,\nonumber\\
\label{KineticEquationBroadened}
\end{eqnarray}
with ${\cal L}$ is defined as
\begin{equation}
{\cal{L}}(\Delta\omega)  = 
\frac{\Gamma_{k12}}{(\Delta\omega)^2 + \Gamma_{k12}^2}.
\label{scriptyL}
\end{equation}
Here, as in section (\ref{B2}) $\Gamma_{k12}$ is the total broadening of each particular resonance, and is given below.

The difference between kinetic equation (\ref{KineticEquation}) and 
the generalized kinetic equation (\ref{KineticEquationBroadened}) is that  the energy conserving delta-functions in Eq.~(\ref{KineticEquation}), $\delta({\omega_{{\bf p}} -\omega_{{{\bf p}_1}} - \omega_{{{\bf p}_2}} })$, was ``broadened''.  The physical motivation for this
broadening is the following: when the resonant kinetic equation is
derived, it is assumed that the amplitude of each plane wave is
constant in time, or, in other words, that the lifetime of single
plane wave is infinite.  The resulting kinetic equation, nevertheless, predicts
that wave amplitude changes. Consequently the   
wave lifetime is finite.  For small level of nonlinearity this 
distinction is not significant, and resonant kinetic equation constitutes a 
self-consistent description. For larger values of nonliterary this is no longer the 
case, and the wave lifetime is finite and amplitude changes need to be taken into account.  
Consequently interactions may not be strictly resonant.
This statement also follows from the Fourier uncertainty principle. 
Waves with varying amplitude can not be represented by a single Fourier component.  
This effect is larger for larger normalized Boltzmann rates. 

If the nonlinear frequency renormalization tends to zero, i.e. $\Gamma_{k12} \to 0$, ${\cal L}$ reduces to the delta function (compare to (\ref{DeltaFunction}): 
$$\lim\limits_{\Gamma_{k12}\to 0} {\cal{L}}(\Delta\omega)  =\pi 
\delta(\Delta\omega).$$
Consequently, in the limit resonant interactions (i.e. no broadening) 
(\ref{KineticEquationBroadened}) reduces to (\ref{KineticEquation}) .

If, on the other hand, one does not take the $\Gamma_{\p\p_1\p_2}\to 0$
limit, then one has to calculate $\Gamma_{\p\p_1\p_2}$
self-consistently. To achieve this we realize that by deriving the
generalized kinetic equation (\ref{KineticEquationBroadened}) we 
allow changes in wave amplitude. The rate of change can
be identified from equation (\ref{KineticEquationBroadened}) in the
following way. Let us go through (\ref{KineticEquationBroadened}) term
by term, and identify all term that multiply the $n_\p$ on the right-hand-side.
Those terms can be loosely interpreted as a nonlinear wave damping 
acting on the given wavenumber:
\begin{eqnarray}
\gamma_{{\bf p}} = 4 \int
 |V_{{\bf p}_1,{\bf p}_2}^{{\bf p}}|^2 \, (n_{{{\bf p}}_1}+
n_{{{\bf p}}_2}) \,
\delta_{{{\bf p} - {\bf p}_1-{\bf p}_2}} \, {\cal L}({\omega_{{\bf p}}
-\omega_{{{\bf p}_1}}-\omega_{{{\bf p}_2}}})
d {\bf p}_{12}
\nonumber \\
-4\int
 \, |V_{{\bf p}_2,{\bf p}}^{{\bf p}_1}|^2\, (n_{{{\bf p}}_2}  - n_{{{\bf p}}_1}) 
\, \delta_{{{\bf p}_1 - {\bf p}_2-{\bf p}}} \,
  {\cal L} ({{\omega_{{\bf p}_1} -\omega_{{\bf p}_2}-\omega_{{\bf p}}}})
\, d {\bf p}_{12}
\nonumber \\
-4\int
 \, |V_{{\bf p},{\bf p}_1}^{{\bf p}_2}|^2\, (n_{{{\bf p}}_1}- n_{{{\bf p}}_2})
 \, \delta_{{{\bf p}_2 - {\bf p}-{\bf p}_1}} \,
  {\cal L}({{\omega_{{\bf p}_2} -\omega_{{\bf p}}-\omega_{{\bf p}_1}}})
\, d {\bf p}_{12}
 \ . \nonumber\\
\label{Gamma}
\end{eqnarray}
The interpretation of this formula is the following: nonlinear 
wave-wave interactions lead to the change of wave amplitude, which 
in turn makes the lifetime of the waves to be finite. This, in turn, makes 
the interactions to be near-resonant. 

The next question is how to relate the individual wave damping 
$\gamma_\p$ with the overall broadening of the resonances of three 
interacting waves.  As we have rigorously shown in (\cite{LLNZ}) the errors
add up, so that 
\begin{equation}
\Gamma_{k12}=\gamma_{{\bf p}}+\gamma_{{\bf p}_1}+\gamma_{{\bf p}_2}.
\label{Gammak12}
\end{equation}
It means that the total resonance broadening is the sum of individual
frequency broadening, and can be thus seen as the ``triad
interaction'' frequency. 

A rigorous derivation of the kinetic equation with a broadened delta function is given in details for a general three-wave Hamiltonian system in \citep{LLNZ}.  The derivation is based upon the Wyld diagrammatic technique for non-equilibrium wave systems and utilizes the Dyson-Wyld line resummation.  This resummation permits an analytical resummation of the infinite series of reducible diagrams for Greens functions and double correlators.  Consequently, the resulting kinetic equation is not limited to the Direct Interaction Approximation (DIA), but also includes higher order effects coming from infinite diagrammatic series.  We emphasize, however, that the approach {\em is} perturbative in nature and there are neglected parts of the infinite diagrammatic series.  The reader is referred to \cite{LLNZ} for details of that derivation.  The resulting formulas are given by 
(\ref{KineticEquationBroadened})-(\ref{Gammak12}).

A self-consistent estimate of $\gamma_{{\bf p}} $ requires an iterative solution of (\ref{KineticEquationBroadened}) and
(\ref{Gamma}) over the entire field: the width of the resonance
(\ref{Gamma}) depends on the lifetime of an individual wave [from
(\ref{KineticEquationBroadened})], which in turn depends on the width
of the resonance (\ref{Gammak12}).  This numerically intensive
computation is beyond the scope of this manuscript.  Instead, we make
the uncontrolled approximation that:
\begin{equation}
\gamma_{{\bf p}} = \delta \omega_{{\bf p}}.
\label{GammaFraction}
\end{equation}

We note that this 
choice is made for illustration purposes only, we certainly do not claim that 
it represents a self consistent choice. Below, we will take $\delta$ to be $10^{-3}$ and $10^{-2}$ and $10^{-1}$. These values are rather small, therefore we
remain in the closest proximity to the resonant interactions. 
To show the effect of strong resonant manifold smearing we also investigate
the case with $\delta = 0.5$.

We note in passing that the near-resonant interactions
of the waves were also considered in the \citep{JansenXXX}. There,
instead of our ${\cal{L}}(x)$ function, given by (\ref{scriptyL}), the
corresponding function was given by $\sin({ \pi x})/x$. We have shown in \cite{K03}
that the resulting kinetic equation does {\it not} retain positive definite values of wave action.
To get around that difficulty, self-consistent formula for broadening
or rigorous diagrammatic resummation should be used.

\subsection{Numerical Methods}

Estimates of near-resonant transfers are obtained by assuming
horizontal isotropy and integrating (\ref{KineticEquationBroadened}) over
horizontal azimuth:
\begin{eqnarray}
\frac{\partial n_{{\bf p}}}{\partial t} = 4\pi \int
\frac{k_1 k_2}{S_{p12}} |V_{{\bf p}_1,{\bf p}_2}^{{\bf p}}|^2 \, f_{p12} \,
\delta_{{{\bf p} - {\bf p}_1-{\bf p}_2}} \, {\cal L}({\omega_{{\bf p}}
-\omega_{{{\bf p}_1}}-\omega_{{{\bf p}_2}}})
dk_{12} dm_1
\nonumber \\
-4\pi \int
 \, \frac{k_1 k_2}{S_{12p}} |V_{{\bf p}_2,{\bf p}}^{{\bf p}_1}|^2\,
f_{12p}\, \delta_{{{\bf p}_1 - {\bf p}_2-{\bf p}}} \,
 {\cal L} ({{\omega_{{\bf p}_1} -\omega_{{\bf p}_2}-\omega_{{\bf p}}}})
\, dk_{12} dm_1
\nonumber \\
-4\pi \int
 \, \frac{k_1 k_2}{S_{2p1}} |V_{{\bf p},{\bf p}_1}^{{\bf p}_2}|^2\,
f_{2p1}\, \delta_{{{\bf p}_2 - {\bf p}-{\bf p}_1}} \,
 {\cal L}({{\omega_{{\bf p}_2} -\omega_{{\bf p}}-\omega_{{\bf p}_1}}})
\, dk_{12} dm_1
\, ,
\label{IntKineticEquationBroadened}
\end{eqnarray}
where $S_{p12}$ is the area of the triangle ${\bf k} = {\bf k}_1 + {\bf k}_2$.
We numerically integrated (\ref{IntKineticEquationBroadened})
for ${\bf p}$'s which have frequencies from $f$ to $N$ 
and vertical wavenumbers from $2\pi/(2b)$ to $260\pi/(2b)$.  
The limits of integration are restricted by horizontal wavenumbers from $2\pi/10^5$ to $2\pi/5$ meters$^{-1}$, vertical wavenumbers from $2\pi/(2b)$ to $2\pi/5$ meters$^{-1}$, and frequencies from $f$ to $N$.  The integrals over $k_1$ and $k_2$
are obtained in the kinematic box in $k_1-k_2$ space.  The grids in the {$k_1-k_2$} domain have $2^{17}$ points that  are distributed heavily around the corner of the kinematic box.  The integral over $m_1$ is obtained with $2^{13}$ grid points, which are also distributed heavily for the small vertical wavenumbers whose absolute values are less than $5m$, where $m$ is the vertical wavenumber. 

To estimate the normalized Boltzmann rate we need to choose a form of spectral energy density of internal waves. We utilize the Garrett and Munk spectrum as an agreed-upon representation of the internal waves:  
\begin{equation}
E(\omega,m) = \frac{4 f}{\pi^2 m_{\ast}}  E_0 \frac{1}{1+(\frac{m}{m_{\ast}})^2 } 
\frac{1}{\omega \sqrt{\omega^2-f^2} }  ~ .
\label{GM}
\end{equation}
Here the reference wavenumber is given by 
\begin{equation} 
m_{\ast} = \pi j_{\ast} / b,
\label{Jstar}
\end{equation} 
in which the variable $j$ represents the vertical mode number of an ocean with
an exponential buoyancy frequency profile having a scale height of $b$.

We choose the following set of parameters:
\begin{itemize}
\item $b$ = 1300 m in the GM model
\item The total energy is set as: 
\begin{equation}
E_0 = 30 \times 10^{-4} {\rm ~ m}^2 {\rm ~ s}^{-2} . \nonumber
\end{equation}
\item Inertial frequency is given by $f=10^{-4}$rad/s, and buoyancy frequency is given by $N_0=5 \times 10^{-3}$rad/s. 
\item The reference density is taken to be 
$\rho_0=10^{3}$kg/m$^{3}$.
\item A roll-off corresponding to $j_{\ast}=3$.  
\end{itemize}

We then calculate the normalized Boltzmann rate (\ref{NonlinearRatio}) using  four values of $\delta$ in (\ref{GammaFraction}): $\delta=10^{-3}$, $\delta=10^{-2}$,
$\delta = 10^{-1}$ and $\delta=0.5$. 

\section{Time Scales}
\subsection{Resonant Interactions}\label{resonant_time}

Here we present evaluations of the \cite{LT2} kinetic equation.  These estimates differ from evaluations presented in \cite{O76, McComas, MM81b, PMW80} in that the numerical algorithm includes a finite breadth to the resonance surface whereas previous evaluations have been {\em exactly} resonant.  Results discussed in this section are as close to resonant as we can make ($\delta = 1\times 10^{-3}$).  

We see that for small vertical wavenumbers the normalized Boltzmann rate is of the order of tenth of the wave period. This can be argued to be relatively within the domain of weak nonlinearity. However for increased wavenumbers the level of nonlinearity increases and reaches the level of wave-period (red, or dark blue). There is also  a white region indicating values smaller than minus one. 

We also define a ``zero curve'' - It is the locus of wavenumber-frequency where the normalized Boltzmann rate and time-derivative of waveaction is exactly zero.  The zero curve clearly delineates a pattern of energy gain for frequencies $f < \omega < 2f$, energy loss for frequencies $2f < \omega < 5 f$ and energy gain for frequencies $5f < \omega < N$.  We interpret the relatively sharp boundary between energy gain and energy loss across $\omega = 2f$ as being related to the Parameteric Subharmonic Instability and the transition from energy loss to energy gain at $\omega = 5f$ as a transition from energy loss associated with the Parametric Subharmonic Instability to energy gain associated with the Elastic Scattering mechanism.  See Section \ref{discussion} for further details about this high frequency interpretation.  

The $O(1)$ normalized Boltzmann rates at high vertical wavenumber are surprising given the substantial literature that regards the GM spectrum as a stationary state.  We do not believe this to be an artifact of the numerical scheme for the following reasons.  First, numerical evaluations of the integrand conserve energy to within numerical precision as the resonance surface is approached, consistent with energy conservation property associated with the frequency delta function.  Second, the time scales converge as the resonant width is reduced, as demonstrated by the minimal difference in time scales using $\delta = 1\times 10^{-3} {\rm ~and~} 1\times10^{-2}$.  Third, our results are consistent with approximate analytic expressions (e.g. \cite{MM81a}) for the Boltzmann rate.  Finally, in view of the differences in the representation of the wavefield, numerical codes and display of results, we interpret our resonant ($\delta=0.001$) results as being consistent with numerical evaluations of the resonant kinetic equations presented in \cite{O76, McComas, MM81b, PMW80}.  

As a quantitative example, consider estimates of the time rate of change of low-mode energy appearing in Table 1 of \cite{PMW80}, repeated as row 3 of our Table \ref{Edot} \footnote{A potential interpretation is that this net energy flow out of the non-equilibrium part of the spectrum represents the energy requirements to maintain the spectrum.  }.  We find agreement to better than a factor of two.  In order to explain the remaining differences, you have to examine the details:  \cite{PMW80} use a Coriolis frequency corresponding to $30^{\circ}$ latitude, neglect internal waves having horizontal wavelengths greater than 100 km (same as here) and exclude frequencies $\omega > N_o/3$, with $N_o = 3$ cph.  We include frequencies $f < \omega < N_o$ with Coriolis frequency corresponding to $45^{\circ}$ latitude.  Of possible significance is that \cite{PMW80} use a vertical mode decomposition with exponential stratification with scale height $b=1200$ m (we use $b=1300$ m).  Table \ref{Edot} presents estimates of the energy transfer rate by taking the depth integrated transfer rates of \cite{PMW80}, assuming $\dot{E} \propto N^2$ and normalizing to $N=$ 3 cph.  While this accounts for the nominal buoyancy scaling of the energy transport rate, it does {\em not} account for variations in the distribution of $\dot{E}(m)$ associated with variations in $N$ via $m_{\ast} = \frac{N}{N_o} j_{\ast}/b$ in their model.  Finally, their estimates of $\dot{E}(m)$ are arrived at by integrating only over regions of the spectral domain for which $\dot{E}(m,\omega)$ is negative.  


\subsection{Near-Resonant Interactions}\label{near_resonant_time}

Substantial motivation for this work is the question of whether the GM76 spectrum represents a stationary state.  We have seen that numerical evaluations of a resonant kinetic equation return $O(1)$ normalized Boltzmann rates and hence we are lead to conclude that GM76 is {\em not} a stationary state with respect to resonant interactions.  But the inclusion of near-resonant interactions could alter this judgement.  

Our investigation of this question is currently limited by the absence of an iterative solution to (\ref{KineticEquationBroadened}) and (\ref{Gamma}) and consequent choice to parameterize the resonance broadening in terms of (\ref{GammaFraction}).  However, as we go from nearly resonant evaluations ($10^{-3}$ and $10^{-2}$) to incorporating significant broadening ($10^{-1}$ and 0.5), we find a significant decreases in the normalized Boltzmann rate.  The largest decreases are associated with an expanded region of energy loss associated the Parametric Subharmonic Instability, in which minimum normalized Boltzmann rates change from -3.38 to -0.45 at $(\omega,m b / 2 \pi) = (2.5f, 150)$.  Large decreases here are not surprising given the sharp boundary between regions of loss and gain in the resonant calculations.  Smaller changes are noted within the Induced Diffusion regime.  Maximum normalized Boltzmann rates change from 2.6 to 1.5 at $(\omega,m b / 2 \pi) = (8f, 260)$.  Broadening of the resonances to exceed the boundaries of the spectral domain could be making a contribution to such changes.  

We regard our calculations here as a preliminary step to answering the question of whether the GM76 spectrum represents a stationary state with respect to nonlinear interactions.  Complementary studies could include comparison with analyses of numerical solutions of the equations of motion.  

\section{Discussion}\label{discussion}
\subsection{Resonant Interactions}

\begin{figure*}
\noindent\includegraphics[width=19pc,angle=0]{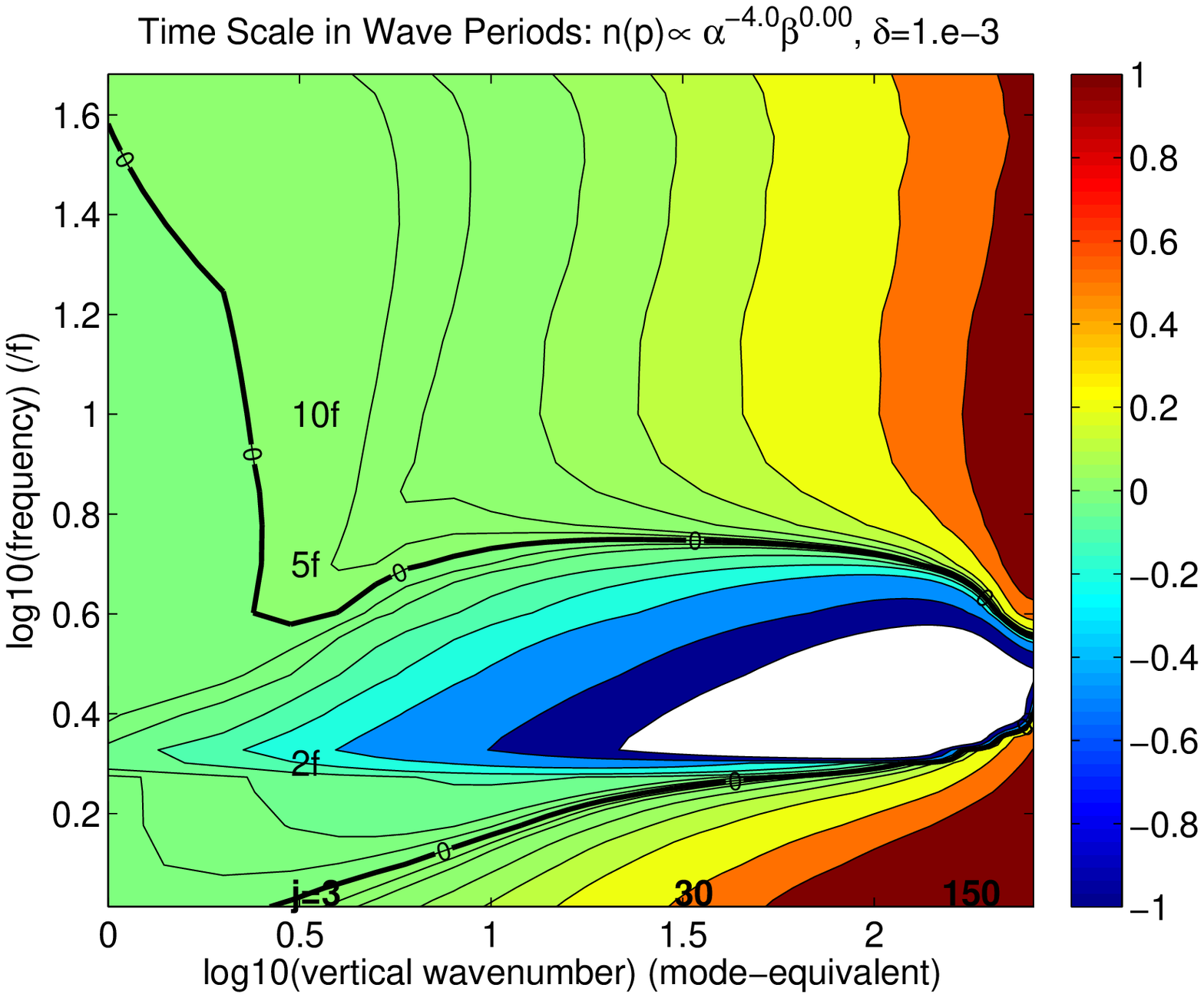} 
\noindent\includegraphics[width=19pc,angle=0]{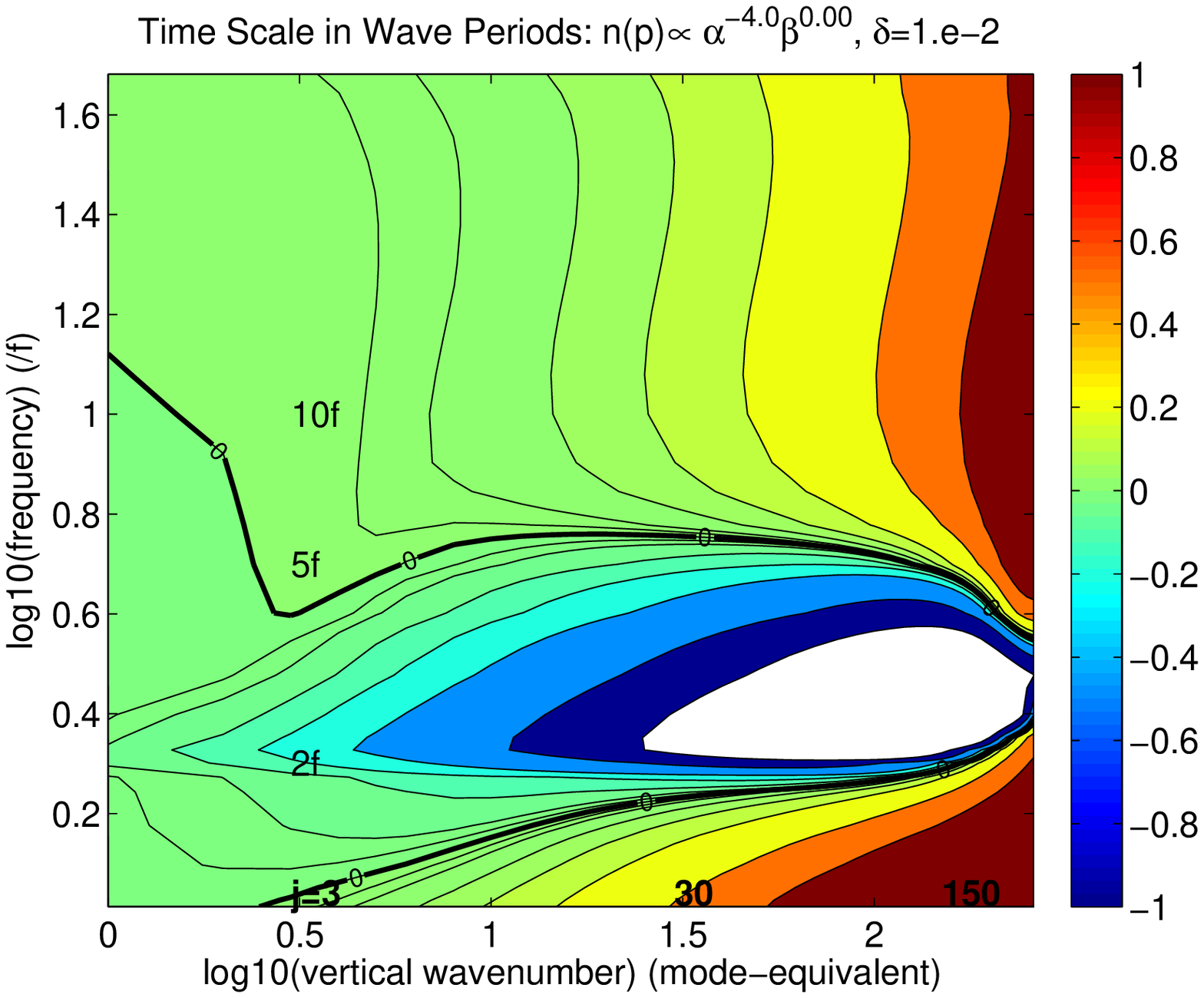} \\
\noindent\includegraphics[width=19pc,angle=0]{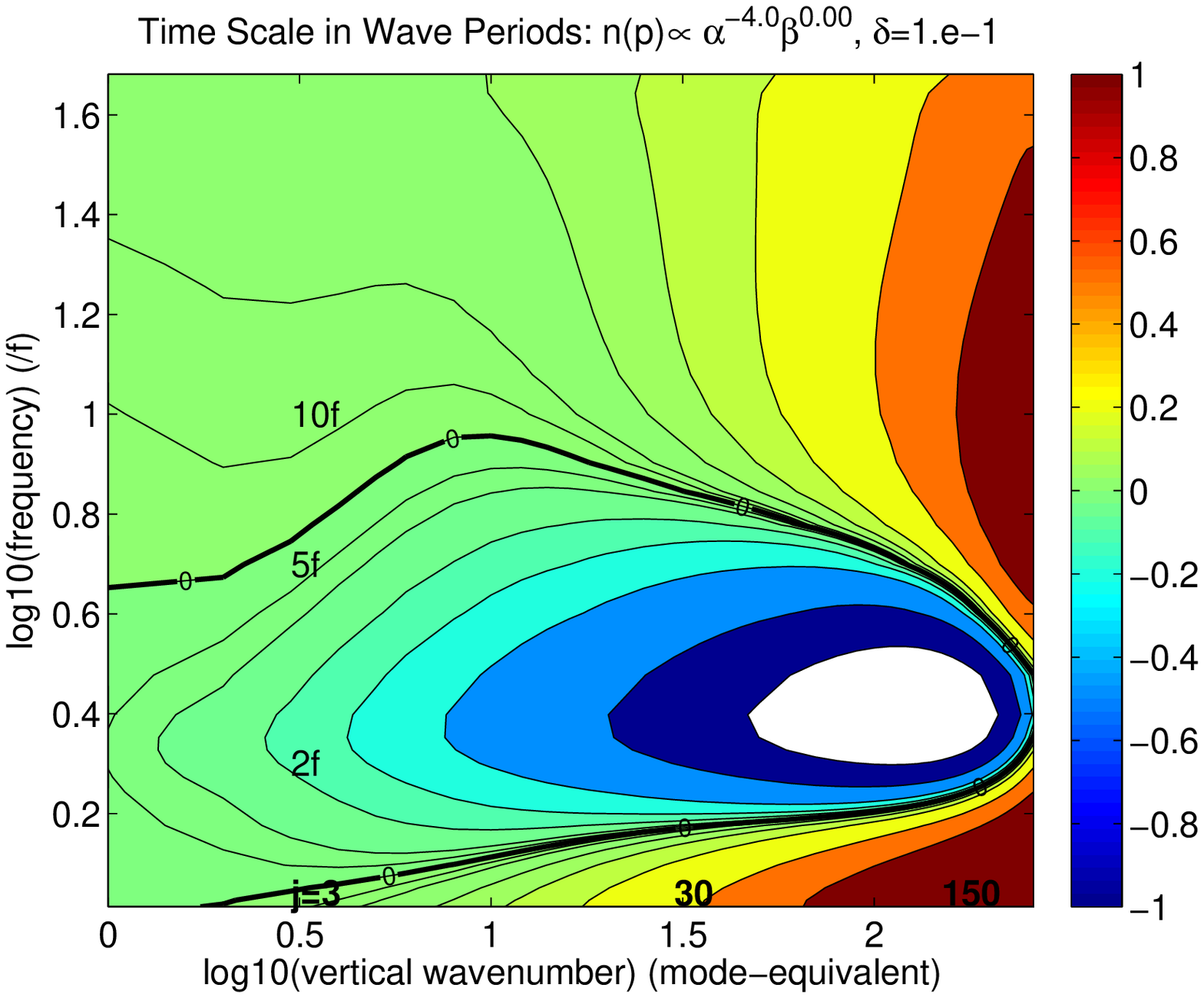} 
\noindent\includegraphics[width=19pc,angle=0]{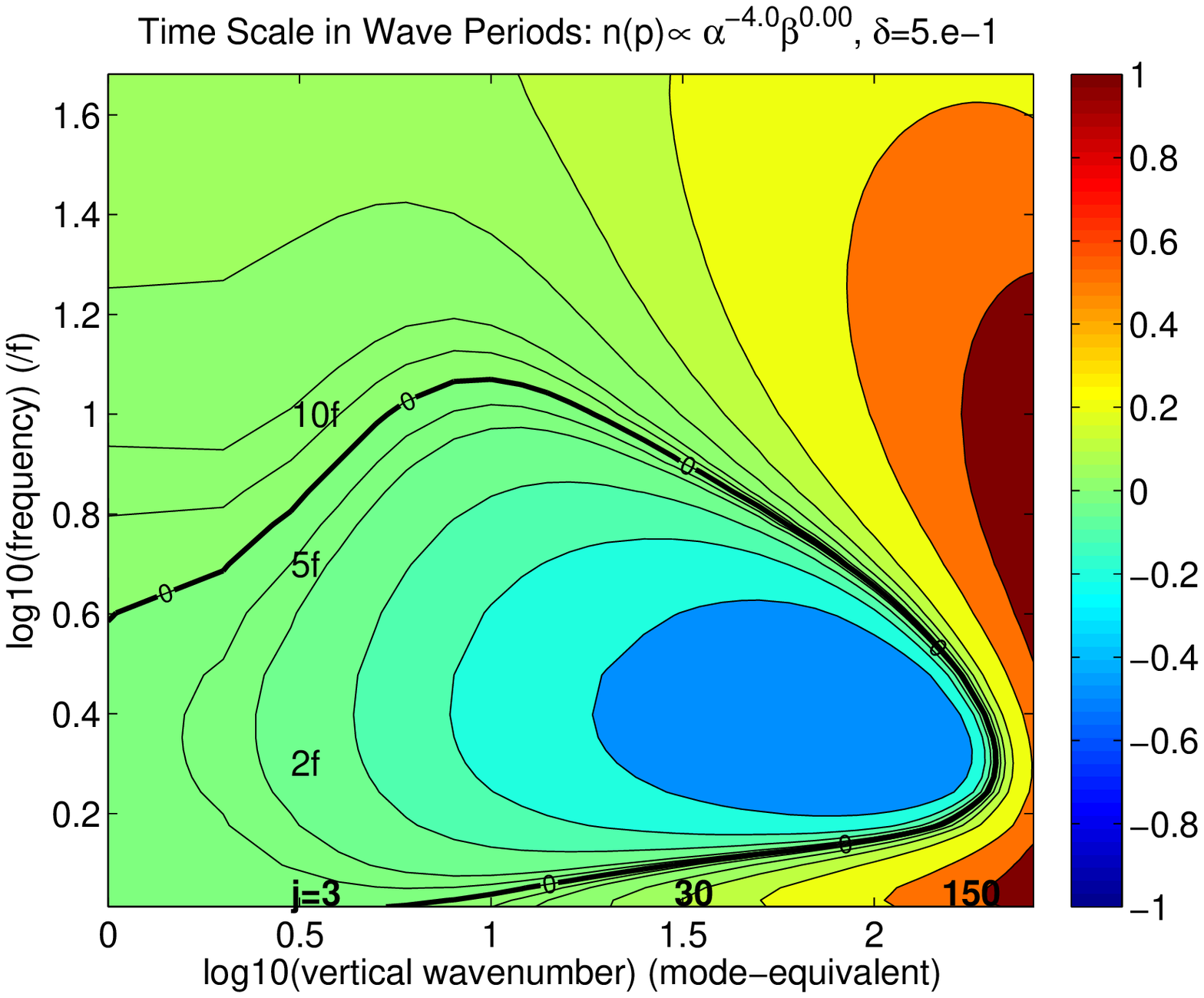} \\
\caption{Normalized Boltzmann rates (\ref{NonlinearRatio}) for the Garrett
and Munk spectrum (\ref{GM}) calculated via (\ref{KineticEquationBroadened}).  
Figures represent normalized Boltzmann rate calculated using
\cite{LT2}, equation (\ref{LTV}) with $\delta=10^{-3}$, (upper-left),
$\delta=10^{-2}$, (upper-right), $\delta=10^{-1}$, (lower-left), $\delta=0.5$, (lower-right). On these 
figures white region on the figure corresponds
to extremely fast time scales, faster then a linear time scale. 
}
\label{NonlinearityParameter} 
\end{figure*}

Several loose ends need to be tied up regarding the assertion that the GM76 spectrum does not constitute a stationary state with respect to resonant interactions.  The first is the interpretation of \cite{MM81b}'s inertial-range theory with constant downscale transfer of energy.  This constant downscale transfer of energy was obtained by patching together the induced diffusion and parametric subharmonic instability mechanisms and is attended by the following caveats:  First, the inertial subrange solution is found only after integrating over the frequency domain and numerical evaluations of the kinetic equation demonstrate that the "inertial subrange" solution also requires dissipation to balance energy gain at high vertical wavenumber.  It takes a good deal of patience to wade through their figures to understand how figures in \cite{MM81b} plots relate to the initial tendency estimates in Figure \ref{NonlinearityParameter}.  Second, \cite{PMW80} argue that GM76 is an near-equilibrium state because of a 1-3 order of magnitude cancellation between the Langevin rates in the induced diffusion regime.  But this is just the $\omega^2/f^2$ difference between the fast and slow induced diffusion time scales.  It does NOT imply small values of the slow induced diffusion time scale, which are equivalent to the normalized Boltzmann rates.  Third, the large normalized Boltzmann rates determined by our numerical procedure are associated with the elastic scattering mechanism rather than induced diffusion.  Normalized Boltzmann rates for the induced diffusion and elastic scattering mechanisms are:
\begin{eqnarray}
\displaystyle \epsilon_{id} & = \frac{\pi^2}{20} \frac{m}{m_c} \frac{m^2}{m^2+ m_{\ast}^2\frac{\omega^2}{f^2} } \nonumber \\ 
\displaystyle \epsilon_{es} & = \frac{\pi^2}{20} \frac{m}{m_c} \frac{m^2}{m^2+ 0.25m_{\ast}^2} \nonumber 
\end{eqnarray}
in which $m_{\ast}$ represents the low wavenumber roll-off of the vertical wavenumber spectrum (vertical mode-3 equivalent here), $m_c$ is the high wavenumber cutoff, nominally at 10 m wavelengths and the GM76 spectrum has been assumed.  The normalized Boltzmann rates for ES and ID are virtually identical at high wavenumber.  They differ only in how their respective triads connect to the $\omega = f$ boundary.  Induced diffusion connects along a curve whose resonance condition is approximately that the high frequency group velocity match the near-inertial vertical phase speed, $\omega/m=f/m_{ni}$.  Elastic scattering connects along a simpler $m = 2m_{ni}$.  Evaluations of the kinetic equation reveal nearly vertical contours throughout the vertical wavenumber domain, consistent with ES, rather than sloped along contours of $\omega \propto m$ emanating from $m=m_{\ast}$ as expected with the ID mechanism.  

The identification of the ES mechanism as being responsible for the large normalized Boltzmann rates at high vertical wavenumber requires further explanation.  The role assigned to the ES mechanism by \cite{MB77} is the equilibration of a vertically anisotropic field.  This can be seen by taking the near-inertial component of a triad to represent ${\bf p}_1$, assuming that the action density of the near-inertial field is much larger than the high frequency fields, and taking the limit $(k,l,m)=(k_2,l_2,-m_2) \equiv {\bf p}^-$.  Thus:
$$
 f_{p12} = n_{\p_1}n_{\p_2} - n_{\p} (n_{\p_1}+n_{\p_2}) \cong n_{\p_1}[n_{\p^{-}} - n_{\p} ]
$$
and transfers proceed until the field is isotropic: $n_{\p^{-}} = n_{\p}$ .  But this is {\bf not} the complete story.  A more precise characterization of the resonance surface takes into account the frequency resonance requiring $\omega - \omega_2 = \omega_1 \cong f$ requires $O(\omega/f)$ differences in $m$ and $-m_2$ if $k=k_2$ and $O(\omega/f)$ differences in $k$ and $k_2$ if $m=-m_2$.  For an isotropic field:
$$
 f_{p12} = n_{\p_1}n_{\p_2} - n_{\p}(n_{\p_1}+n_{\p_2}) \cong n_{p_1}[n_{\p+\delta \p} - n_{\p}] \cong n_{\p_1} [\delta \p \cdot \nabla n_{\p}]
$$
and due care needs to be taken that $\p_1$ is on the resonance surface in the vicinity of the inertial cusp.

\subsection{Near-Resonant Interactions}

The idea of trying to self consistently find the smearing of the
delta-functions is not new.  For internal waves it appears in \cite{DWW82, CF83, DWW84}.  

\cite{DWW82} set up a general framework for a self consistent calculation similar in spirit to \cite{LLNZ}, 
using a path-integral formulation of the diagrammatic technique.
The paper makes an uncontrolled approximation that their nonlinear frequency renormalization 
$\Sigma(\p,\omega)$ is independent of $\omega$, and shows that this
assumption is not self-consistent. \cite{LLNZ} present a more sophisticated approach to a 
self-consistent approximation to
the operator $\Sigma(\p,\omega)$.  In particular, \cite{DWW82} suggests 
$$\Sigma(\p,\omega) = \Sigma(\p,\omega_\p),$$
while \cite{LLNZ} propose a more self-consistent 
$$\Sigma(\p,\omega) = \Sigma[\p,\omega_\p + i \Im \Sigma(\p,\Omega_\p)].$$

\cite{DWW84} evaluate the self-consistency of the resonant interaction
approximation and find that for high-frequency-high-wavenumbers,
the resonant interaction representation is not self-consistent.  A possible critique of these papers is that they use resonant matrix elements given by \cite{MO75} with out appreciating that those elements can only be used strictly on the resonant manifold. 

\cite{CF83} present similar expressions for two-dimensional stratified internal waves. There the kinetic equation is (7.4) with the triple correlation time given by $\Theta$
(our ${\cal L}$) of their (8.7). The key step is to find the
level of smearing of the delta-function, denoted as $\mu_k$ in their
(8.7) (our $\gamma$). This can be achieved by their (8.6), which is similar to our (\ref{Gamma}). The only
difference is that (8.6) hFas slightly different positions of the
poles $i(\gamma_{\bf p_1} + \gamma_{\bf p_2})$, instead of ours
$i(\gamma_{\bf p_1} + \gamma_{\bf p_2}+\gamma_{\bf p})$. Carnavale
points out that the Direct Interaction Approximation leads to his expression, not the sum of
all three $\gamma$'s. We respectfully disagree. However, this is irrelevant for the purpose of this paper, since we do not solve it self consistently anyway, but propose an uncontrolled
approximation (\ref{GammaFraction}). The main advantage of our approach over \cite{CF83} is that we use systematic Hamiltonian structures which are equivalent to the primitive equations of motion, rather than a simplified two-dimensional model. 

\section{Conclusion}\label{Conclusion}

Our fundamental result is that the GM spectrum is {\em not} stationary with respect to the resonant interaction approximation.  This result is contrary to much of the perceived wisdom and gave us cause to review published results concerning resonant internal wave interactions.  We then included near-resonant interactions and found significant reductions in the temporal evolution of the GM spectrum.  

We compared the interaction matrices for three different Hamiltonian formulations and one non-Hamiltonian formulation in the resonant limit.  Two of the Hamiltonian formulations are canonical and one \citep{LT2} avoids a linearization of the Hamiltonian prior to assuming an expansion in terms of weak nonlinearity.  Formulations in Eulerian, isopycnal and Lagrangian coordinate systems were considered.  All four representations lead to {\em equivalent} results on the resonant manifold in the absence of background rotation.  The two representations that include background rotation, a canonical Hamiltonian formulation in isopycnal coordinates and a non-canonical Hamiltonian formulation in Lagrangian coordinates, also lead to {\em equivalent} results on the resonant manifold.  This statement is not trivial given the different assumptions and coordinate systems that have been used for the derivation of the various kinetic equations. It points to an internal consistency on the resonant manifold that we still do not completely understand and appreciate. 

We rationalize the consistent results as being associated with potential vorticity conservation.  In the isopycnal coordinate canonical Hamilton formulation potential vorticity conservation is explicit.  In the Lagrangian coordinate non-canonical Hamiltonian, potential vorticity conservation results from a projection onto the linear modes of the system.  The two non-rotating formulations prohibit relative vorticity variations by casting the velocity as a the gradient of a scalar streamfunction.  

We infer that the non-stationary results for the GM spectrum are related to a higher order approximation of the elastic scattering mechanism than considered in \cite{MB77} and \cite{MM81a}.  

Our numerical results indicate evolution rates of a wave period at high vertical wavenumber, signifying a system which is not weakly nonlinear.  To understand whether such non-weak conditions could give rise to competing effects that render the system stationary, we considered resonance broadening.  We used a kinetic equation with broadened frequency delta function derived for a generalized three-wave Hamiltonian system in \citep{LLNZ}.  The derivation is based upon the Wyld diagrammatic technique for non-equilibrium wave systems and utilizes the Dyson-Wyld line resummation.  This broadened kinetic equation is perceived to be more sophisticated than the two-dimensional direct interaction approximation representation pursued in \cite{CF83} and the self-consistent calculations of 
\cite{DWW84} which utilized the resonant interaction matrix of \cite{O76}.  We find a tendency of resonance broadening to lead to more stationary conditions.  However, our results are limited by an uncontrolled approximation concerning the width of the resonance surface.  

Reductions in the temporal evolution of the internal wave spectrum at high vertical wavenumber were greatest for those frequencies associated with the PSI mechanism, i.e. $f < \omega < 5f$.  Smaller reductions were noted at high frequencies.  

A common theme in the development of a kinetic equation is a perturbation expansion permitting the wave interactions and evolution of the spectrum on a slow time scale, e.g. Section \ref{B2}.  An assumption of Gaussian statistics at zeroth order permits a solution of the first order triple correlations in terms of the zeroth order quadruple correlations.  Assessing the adequacy of this assumption for the zeroth order high frequency wavefield is a challenge for future efforts.  Such departures from Guassianity could have implications for the stationarity at high frequencies.  

Nontrivial aspects of our work are that we utilize the canonical Hamiltonian representation of \cite{LT2} which results in a kinetic equation without first linearizing to obtain interaction coefficients defined only on the resonance surface and that the broadened closure scheme of \cite{LLNZ} is more sophisticated than the Direct Interaction Approximation.  Inclusion of interactions between internal waves and modes of motion associated with zero eigen frequency, i.e. the vortical motion field, is a challenge for future efforts.  

We found no coordinate dependent (i.e. Eulerian, isopycnal or Lagrangian) differences between interaction matrices on the resonant surface.  We regard it as intuitive that there will be coordinate dependent differences off the resonant surface.  It is a robust observational fact that Eulerian frequency spectra at high vertical wavenumber are contaminated by vertical Doppler shifting:  near-inertial frequency energy is Doppler shifted to higher frequency at approximately the same vertical wavelength.  Use of an isopycnal coordinate system considerably reduces this artifact \citep{SandP91}.  Further differences are anticipated in a fully Lagrangian coordinate system \citep{Pinkel08}.  Thus differences in the approaches may represent physical effects and what is a stationary state in one coordinate system may not be a stationary state in another.  Obtaining canonical coordinates in an Eulerian coordinate system with rotation and in the Lagrangian coordinate system are challenges for future efforts.  
\clearpage

\begin{acknowledgment} 
 
We thank V. E. Zakharov for presenting us with a book \citep{Mir} and for encouragement.  We also thank E. N. Pelinovsky for providing us with \cite{PR77}.  We greatfully acknowledge funding provided by a Collaborations in Mathematical Geosciences (CMG) grant from the National Science Foundation.    YL is also supported by NSF DMS grant 0807871 and ONR  Award N00014-09-1-0515.  We are grateful to YITP in Kyoto University for permitting use of their facility.

\end{acknowledgment}
\clearpage


\begin{appendix}
\section*{\begin{center}Matrix Elements\end{center}}\label{VariousApproaches}

Our attention is restricted to the hydrostatic balance case, for which 
\begin{equation}
\mid{\bf k}\mid \ll \mid m \mid \ .
\label{hydrostatic}
\end{equation} 
A minor detail is that the linear frequency has different algebraic representations in isopycnal and  Cartesian coordinates. The Cartesian  vertical wavenumber, $k_z$, and the density wavenumber, $m$, are related as $m = - g/(\rho_0 N^2) k_z$ where $g$ is gravity, $\rho$ is density with reference value $\rho_0$, $N$ is the buoyancy (Brunt--V\"{a}is\"{a}l\"{a}) frequency and $f$ is the Coriolis frequency.  
In isopycnal coordinates the dispersion relation is given by,
\begin{eqnarray}
\omega(\p)  = \sqrt{f^2 + \frac{g^2}{\rho_0^2 N^2} \frac{\mid {\bf k}\mid^2}{m^2}}.
 \label{eq:dispersionISO}
\end{eqnarray}
In Cartesian coordinates, 
\begin{eqnarray}
\omega(\p) = \sqrt{f^2 + N^2 \frac{\mid {\bf k}\mid^2}{k_z^2}}~.
\label{eq:dispersionCAR}
\end{eqnarray}
In the limit of $f=0$ these dispersion relations assume the 
form  
\begin{equation}
\omega_{\p} \propto  \frac{ \mid{\bf k} \mid}{\mid m \mid} \propto \frac{ \mid {\bf k} \mid}{\mid k_z \mid} 
\label{dispersion}
\end{equation}

\subsection*{M\"uller and Olbers}

Matrix elements derived in  \citet{O74} are given by
$\mid {V^{\p}_{\p_1, \p_2}}^{\mathrm{MO}}\mid^2 = T^{+} / (4\pi)$ and
$\mid {V^{\p_1}_{\p_2, \p}}^{\mathrm{MO}}\mid^2 = T^{-} / (4\pi)$.  We extracted $T^{\pm}$ from the Appendix of \citet{MO75}.  In our notation, in the hydrostatic balance approximation, their matrix elements are given by

\begin{align}
|{V^{\p}_{\p_1, \p_2}}^{\mathrm{MO}}|^2 =\frac{(N_0^2-f^2)^2}{32 \rho_0} \omega \omega_1 \omega_2
\left|
 \frac{|{\bf k}| |{\bf k}_1| |{\bf k}_2|}{\omega \omega_1 \omega_2 |{\bf p}||{\bf p}_1||\p_2|}
\right.
\nonumber\\
\left(
 - \frac{\left(-m_1 \frac{{\bf k}_1 \cdot {\bf k}_2 - i f {\bf k}_2 \cdot {\bf k}_1^{\perp}/\omega_1}{k_1^2} + m_2\right) \left(-m_2 \frac{{\bf k}_1 \cdot {\bf k}_2  - i f {\bf k}_1 \cdot {\bf k}_2^{\perp}/\omega_2}{k_2^2} + m_1\right)}{m}
\right.
\nonumber\\
 - \frac{\left(-m_2 \frac{{\bf k}_2 \cdot {\bf k} + i f {\bf k}_2 \cdot {\bf k}^{\perp}/\omega_2}{k_2^2} + m\right)
 \left(-m \frac{{\bf k}_2 \cdot {\bf k}  - i f {\bf k} \cdot {\bf k}_2^{\perp}/\omega}{k^2} + m_2\right)}{m_1}
\nonumber\\
\left.\left.
 - \frac{\left(-m \frac{{\bf k} \cdot {\bf k}_1 - i f {\bf k} \cdot {\bf k}_1^{\perp}/\omega}{k^2} + m_1\right)
 \left(-m_1 \frac{{\bf k} \cdot {\bf k}_1 + i f {\bf k}_1 \cdot {\bf k}^{\perp}/\omega_1}{k_1^2} + m\right)}{m_2}
\right)
\right|^2
.
\label{VMO}
\end{align}

Taking a $f=0$ limit  reduces the problem to scale invariant problem.
We get the following simplified expression:
\begin{eqnarray}
 |{V^{\p}_{\p_1, \p_2}}^{\mathrm{MO}}|^2 \propto
\frac{|{\bf k}||{\bf k}_1||{\bf k}_2|}{|m m_1 m_2|}
 \left(
 - \frac{1}{m}
 \left(-\frac{m_2 {\bf k}_1 \cdot {\bf k}_2}{|{\bf k}_2|^2} + m_1 \right)
 \left(-\frac{m_1 {\bf k}_2 \cdot {\bf k}_1}{|{\bf k}_1|^2} + m_2 \right)
\right.
\nonumber
\\
\left.
 + \frac{1}{m_1}
 \left(\frac{m_2 {\bf k} \cdot {\bf k}_2}{|{\bf k}_2|^2} - m \right)
 \left(-\frac{m {\bf k}_2 \cdot {\bf k}}{|{\bf k}|^2} + m_2 \right)
 + \frac{1}{m_2}
 \left(-\frac{m {\bf k}_1 \cdot {\bf k}}{|{\bf k}|^2} + m_1 \right)
 \left(\frac{m_1 {\bf k} \cdot {\bf k}_1}{|{\bf k}_1|^2} - m \right)
 \right)^2
\end{eqnarray}
This simplified expression is going to be used for comparison of approaches 
in section (\ref{ResonantInteractions}).

\subsection*{Voronovich}
We formulate the matrix elements for Voronovich's Hamiltonian using his formula (A.1).  This formula is derived for general boundary conditions. To compare with other matrix elements of this paper, we assume a constant stratification profile and Fourier basis as the vertical structure function $\phi(z)$. That allows us to solve for the matrix elements defined via Eq.~(11) and above it in his paper.
Then the convolutions of the basis functions give delta-functions in vertical wavenumbers.
Vornovich's equation (A.1) transforms into:
 \begin{eqnarray}
  |{V^{\p}_{\p_1, \p_2}}^{\mathrm{V}}|^2 \propto
 \frac{|{\bf k}||{\bf k}_1||{\bf k}_2|}{|m m_1 m_2|}
 \left(
 - m
 \left(
 \frac{1}{|{\bf k}| |m|}
 \left(\frac{{\bf k} \cdot {\bf k}_1 |m_1|}{|{\bf k}_1|} + \frac{{\bf k}
 \cdot {\bf k}_2 |m_2|}{|{\bf k}_2|} \right)
 + \frac{\omega_1 + \omega_2 - \omega}{\omega}
 \right)
 \right.
 \nonumber\\
 \left.
 + m_1
 \left(
 \frac{1}{|{\bf k}_1| |m_1|}
 \left(\frac{{\bf k} \cdot {\bf k}_1 |m|}{|{\bf k}|} + \frac{{\bf k}_1
 \cdot {\bf k}_2 |m_2|}{|{\bf k}_2|} \right)
 - \frac{\omega_1 + \omega_2 - \omega}{\omega_1}
 \right)
 \right.
 \nonumber\\
 \left.
 + m_2
 \left(
 \frac{1}{|{\bf k}_2| |m_2|}
 \left(\frac{{\bf k} \cdot {\bf k}_2 |m|}{|{\bf k}|} + \frac{{\bf k}_2
 \cdot {\bf k}_1 |m_1|}{|{\bf k}_1|} \right)
 - \frac{\omega_1 + \omega_2 - \omega}{\omega_2}
 \right)
 \right)^2 . \nonumber \\
 \label{eq:Voronovich}
 \end{eqnarray}

Note that Eq.~(\ref{eq:Voronovich}) shares structural similarities with the interaction matrix elements in {\em isopycnal\/} coordinates, Eq.~(\ref{Hamiltonian}) below.

\subsection*{Caillol and Zeitlin}
A non-Hamiltonian kinetic equation for internal waves was derived in
\citet{Zeitlin}, Eq.~(61) directly from the dynamical equations of
motion, without the use of the Hamiltonian structure. 

To make it appear equivalent to more
traditional form of kinetic equation, as in \citet{ZLF}, we make
a change of variables ${\bf l}\to -{\bf l}$ in the second line, and
${\bf k}\to -{\bf k}$ in the third line of (61) of \citet{Zeitlin}.  If
we further assume that all spectra are symmetric, $n(-{\bf p}) =
n({\bf p})$, then the kinetic equation assumes traditional form, as in Eq.~(\ref{KineticEquation}), see
\citet{MO75,ZLF,LT,LT2}.

The matrix elements according to \citet{Zeitlin} are shown as
$X_{k,l,p}$ and $Y_{k,l,p}^{\pm}$ in Eqs.~(62) and (63), where
$|{V^{{\bf p}}_{{\bf p}_1, {\bf p}_2}}^{\mathrm{CZ}}|^2 = X_{{\bf p}_1,{\bf p}_2,{\bf p}}$ and
$|{V^{{\bf p}_1}_{{\bf p}_2, {\bf p}}}^{\mathrm{CZ}}|^2 = Y_{{\bf p}_1,-{\bf p}_2,{\bf p}}^{+}$.
In our notation it reads
\begin{eqnarray}
 |{V^{{\bf p}}_{{\bf p}_1, {\bf p}_2}}^{\mathrm{CZ}}|^2 \propto
(|{\bf k}| \mathrm{sgn}(m) + |{\bf k}_1| \mathrm{sgn}(m_1) + |{\bf k}_2| \mathrm{sgn}(m_2))^2
\frac{(m^2 - m_1 m_2)^2}{|m| |m_1| |m_2| |{\bf k}||{\bf k}_1||{\bf k}_2|}
\nonumber
\\ 
\times\left(
\frac{|{\bf k}|^2 - |{\bf k}_1| \mathrm{sgn}(m_1) |{\bf k}_2| \mathrm{sgn}(m_2)}{m^2 - m_1 m_2} m
- \frac{|{\bf k}_1|^2}{m_1}
- \frac{|{\bf k}_2|^2}{m_2}
\right)^2 \, \nonumber \\.\label{eq:VCZ}
\end{eqnarray}
This expression is going to be used for comparison of approaches 
in section (\ref{ResonantInteractions}). 

\subsection*{Isopycnal Hamiltonian}

Finally, in \citet{LT2} the following wave-wave interaction matrix
element was derived based on a canonical Hamiltonian formulation in isopycnal coordinates:

\begin{align}
 |{V^0_{1,2}}   ^{\mathrm{H}}
|^2 = \frac{N^2}{32 g}
\left(
\left(
\frac{k {\bf k}_1 \cdot {\bf k}_2}{k_1 k_2} \sqrt{\frac{\omega_1 \omega_2}{\omega}}
+ \frac{k_1 {\bf k}_2 \cdot {\bf k}}{k_2 k} \sqrt{\frac{\omega_2 \omega}{\omega_1}}
+ \frac{k_2 {\bf k} \cdot {\bf k}_1}{k k_1} \sqrt{\frac{\omega \omega_1}{\omega_2}}
\right.
\right.
\nonumber\\
\left.
\left.
+ \frac{f^2}{\sqrt{\omega \omega_1 \omega_2}}
\frac{k_1^2 {\bf k}_2 \cdot {\bf k} - k_2^2 {\bf k} \cdot {\bf k}_1 - k^2 {\bf k}_1 \cdot {\bf k}_2}{k k_1 k_2}
\right)^2
\right.
\nonumber\\
\left.
+
\left(
f \frac{{\bf k}_1 \cdot {\bf k}_2^{\perp}}{k k_1 k_2}
 \left(\sqrt{\frac{\omega}{\omega_1 \omega_2}} (k_1^2 - k_2^2)
 - \sqrt{\frac{\omega_1}{\omega_2 \omega}}  (k_2^2-k^2)
 - \sqrt{\frac{\omega_2}{\omega \omega_1}} (k^2-k_1^2)\right)
\right)^2
\right)~\ .\nonumber \\ \label{LTV}
\end{align}
\citet{LT} is a rotationless limit of \citet{LT2}.
Taking the $f\to 0$ limit, the \cite{LT2} reduces to \cite{LT}, and (\ref{LTV}) reduces to 
\begin{eqnarray}
|{V^{{\bf p}}_{{\bf p}_1, {\bf p}_2}}^{\mathrm{H}}|^2 \propto
\frac{1}{|{\bf k}||{\bf k}_1||{\bf k}_2|} \left(
 |{\bf k}| {\bf k}_1 \cdot {\bf k}_2 \sqrt{\left|\frac{m}{m_1 m_2}\right|}
+ |{\bf k}_1| {\bf k}_2 \cdot {\bf k} \sqrt{\left|\frac{m_1}{m_2 m}\right|}
+ |{\bf k}_2| {\bf k} \cdot {\bf k}_1 \sqrt{\left|\frac{m_2}{m m_1}\right|}
 \right)^2 .\nonumber\\
\label{Hamiltonian}
\end{eqnarray}

Observe that in this form, these equations share structural similarities with 
Eq.~(\ref{eq:Voronovich}).  

\end{appendix}
\clearpage

\bibliographystyle{./ametsoc}
\bibliography{./bibliography/references}

\clearpage

\begin{table}[t]
\label{TABLEOFELEMENTS2}
\caption{ A list of various kinetic equations.  
Results from \cite{O76}, \cite{MB77} and \cite{PMW80} are reviewed in \cite{M86}, who state that \cite{O76}, \cite{MB77} and an unspecified Eulerian representation are consistent on the resonant manifold.  \cite{PMW80} utilizes Langevin techniques to assess nonlinear transports.  \cite{M86} characterizes those Langevin results as being mutually consistent with the direct evaluations of kinetic equations presented in \cite{O76} and\cite{MB77}.  \cite{K68} states (without detail) that \cite{K66} and \cite{H66} give numerically similar results.  A formulation in terms of discrete modes will typically permit an arbitrary buoyancy profile, but obtaining results requires specification of the profile.  Of the discrete formulations, \cite{PMW80} use an exponential profile and the others assume a constant stratification rate.  The kinetic equations marked by $^{\dag}$  are investigated in Section \ref{ResonantInteractions}, while kinetic equations marked by $^{\ddag}$ are investigated further in Section \ref{OffResonant}. }
\begin{center}
\begin{tabular}{cccccc}
\hline
source & coordinate & vertical & rotation & hydro- & special \\
 & system & structure & & static & \\
\hline
\citet{H66} & Lagrangian & discrete & no & no & \\
\citet{K66, K68} & Eulerian & discrete & no & no & non-Hamiltonian \\
\citet{MO75}$^{\dag \ddag}$ & Lagrangian & cont. & yes & no &\\
\citet{McC75, McComas} & Lagrangian & cont. & yes & yes & \\
\citet{PR77} & Lagrangian & cont. & no & no &  Clebsch\\
\citet{Voronovich}$^{\dag}$ & Eulerian & cont. & no & yes & Clebsch \\
\citet{PMW80} & Lagrangian & discrete & yes & no & Langevin \\
\citet{Milder} & Isopycnal & n/a & no & no &  \\
\citet{Zeitlin}$^{\dag}$ & Eulerian & cont. & no & no & non-Hamiltonian \\
\citet{LT}$^{\dag}$ & Isopycnal & cont. & no & yes & canonical \\
\citet{LT2}$^{\ddag}$ & Isopycnal & cont. & yes & yes & canonical \\
\hline
\end{tabular}
\end{center}
\end{table}

\begin{table}[t]
\label{Edot}
\caption{ Numerical evaluations of $ \int_f^N E(m,\omega) d\omega$ for vertical mode numbers 1-8.  The sum is given in the right-most column.  }
\begin{center}
\begin{tabular}{|c|c|c|c|c|c|c|c|c|c|c|}
\hline \\
$\dot{E}\times10^{-10}$ W/kg & & mode-1 & 2 & 3 & 4 & 5 & 6 & 7 & 8 & $\Sigma$ \\
\hline \\
\cite{LT2} & GM76 & -1.46 & -1.72 & -1.76	 & -1.69 & -1.57	 & -1.40 & -1.08	 & -0.81 & -11.5 \\
\hline \\
\cite{PMW80} & GM76 & -1.83 & -2.17 & -2.17 & -1.83 & -1.67 & -1.00 & & & -10.7\\
\hline
\end{tabular}
\end{center}
\end{table}

\end{document}